\documentclass[letterpaper,twocolumn,10pt,anonymous]{article}
\usepackage{usenix}

\usepackage{tikz}
\usepackage{verbatim}
\usepackage{makecell}
\usepackage{pifont}
\usepackage{subfigure}
\usepackage{listings}
\usepackage{booktabs}
\usepackage{diagbox}
\usepackage{enumitem}
\usepackage{balance}
\usepackage{xspace}
\usepackage{tabularx}
\usepackage{framed}
\usepackage{hyperref}
\usepackage{algorithm}
\usepackage{amsmath}
\usepackage{amsfonts}
\usepackage{amsthm}
\usepackage{amssymb}
\usepackage[noend]{algpseudocode}
\usepackage{threeparttable}
\usepackage{enumerate}
\usepackage{balance}

\newcommand{\parabf}[1]{\medskip\noindent\textbf{#1}}
\newcommand{\sysname}{5\-G\-C\-$^2$\-a\-c\-h\-e\xspace}

\algnewcommand{\LineComment}[1]{\Statex \(\) #1}
\algnewcommand\algorithmicforeach{\textbf{foreach}}
\algdef{S}[FOR]{ForEach}[1]{\algorithmicforeach\ #1\ \algorithmicdo}

\newcommand{\squishlist}{
   \begin{list}{$\bullet$}
    { \setlength{\itemsep}{0pt}      \setlength{\parsep}{3pt}
      \setlength{\topsep}{3pt}       \setlength{\partopsep}{0pt}
      \setlength{\leftmargin}{3.5mm} \setlength{\labelwidth}{1em}
      \setlength{\labelsep}{0.5em} } }
      
\newcounter{boxlblcounter}  
\newcommand{\squishnumlist}{
  \begin{list}{\arabic{boxlblcounter}.}
    { \usecounter{boxlblcounter} 
      \setlength{\itemsep}{0pt}      \setlength{\parsep}{3pt}
      \setlength{\topsep}{3pt}       \setlength{\partopsep}{0pt}
      \setlength{\leftmargin}{3.5mm} \setlength{\labelwidth}{1em}
      \setlength{\labelsep}{0.5em} } }

\newcommand{\squishend}{
    \end{list}  }

\newcommand\ie{{\em i.e.,}\xspace}
\newcommand\eg{{\em e.g.,}\xspace}
\newcommand\etc{{\em etc.}\xspace}
\newcommand\etal{{\em et al.}\xspace}

\newcommand{\COMMENTS}{no}

\usepackage{color,balance,xspace,verbatim,ifthen,engord}

\usepackage{gensymb,amsmath,wasysym,amsthm,marvosym}

\usepackage{booktabs,colortbl,diagbox,multirow,tabularx,tablefootnote} 

\usepackage{dblfloatfix} 

\usepackage{algorithm}

\usepackage{url}

\def\eg{\textit{e.g.,}\hspace{1mm}}
\def\ie{\textit{i.e.,}\hspace{1mm}}
\def\etal{\textit{et al.}\hspace{1mm}}
\def\etc{\textit{etc.}\hspace{1mm}}

\usepackage{courier} 

\ifthenelse{\equal{\COMMENTS}{yes}}{
    \newcommand{\yuxiang}[1]{{\color{blue}(Lin: #1)}}
    \newcommand{\weiguo}[1]{{\color{orange}(Wang: #1)}}
    \newcommand{\neil}[1]{{\color{green}(Nie: #1)}}
}{
    \newcommand{\yuxiang}[1]{}
  \newcommand{\neil}[1]{}
  \newcommand{\weiguo}[1]{}
}

\ifthenelse{\equal{\COMMENTS}{yes}}{
  \newcommand{\todo}[1]{\textcolor{red}{\textbf{TODO:} #1}}
  \newcommand{\fye}[1]{\textcolor{red}{#1}}  
  \newcommand{\remind}[1]{\footnote{\textit{\textcolor{red}{\textbf{Remind:} #1}}}}
  
  \newcommand{\del}[1]{\color{blue} {\sout{#1}}}
  \newcommand{\p}[1]{\vskip 1ex \noindent\colorbox{yellow}{\parbox{\columnwidth}{#1}}\vskip 4pt}
  \newcommand{\note}[1]{\vskip 4ex \noindent\colorbox{yellow}{\parbox{\columnwidth}{#1}}\vskip 6ex} 
  \newcommand{\q}[1]{\vskip 1ex \noindent\colorbox{magenta}{\parbox{\columnwidth}{\textbf{Question:} #1}}\vskip 4pt} 
  \newcommand{\qa}[1]{\hl{\textbf{Answer:} #1}}
}{
  \newcommand{\todo}[1]{}
  
  \newcommand{\fye}[1]{}
  \newcommand{\remind}[1]{}

  \newcommand{\del}[1]{}
  \newcommand{\p}[1]{}
  \newcommand{\note}[1]{}

  \newcommand{\q}[1]{}
  \newcommand{\qa}[1]{}  
}

\newcommand{\circled}[1]{\raisebox{.5pt}{\textcircled{\raisebox{-.9pt} {#1}}}}

\newcommand{\wm}[1]{{\color{black}#1}}

\newcommand{\be}[1]{{\color{black}#1}}

\newcommand{\hn}[1]{{\color{black}#1}}

\begin{document}

\date{}

\author{Haonan Jia, Meng Wang, Biyi Li, Yirui Liu, Junchen Guo, Pengyu Zhang\\\{jia.michael.haonan, mengwangwm, juncguo\}@gmail.com\\Alibaba Group}

\title{\sysname: Improving 5G UPF Performance via Cache Optimization}

\maketitle

\begin{abstract}
Last Level Cache (LLC) is a precious and critical resource that impacts the performance of applications running on top of CPUs. In this paper, we reveal the significant impact of LLC on the performance of the 5G user plane \wm{function} (UPF) when running a cloudified 5G core on general-purposed servers. With extensive measurements showing that the throughput can degrade by over 50\% when the precious LLC resource of UPF is not properly allocated, we identify three categories of performance degradation caused by incorrect LLC usage: DMA leakage problem, hot/cold mbuf problem and cache contention. To address these problems, we introduce the design and implementation of \sysname that monitors the LLC status as well as the throughput performance and dynamically adjusts key parameters of the LLC resource allocation. Our experiments show that \sysname enables a commercial 5G core to increase its throughput to 76.41Gbps, 39.41\% higher than the original performance and 29.55\% higher than the state-of-the-art. 
\end{abstract}


\section{Introduction}
\label{sec:intro}
Last Level Cache (LLC) is a precious and critical resource that is shared by many cores of a CPU \cite{llc-background1,llc-background2}. In a 5G core network, there are two types of entities that heavily use LLC for achieving high throughput and low latency: Direct I/O technology (DDIO) \cite{ddio} and 5G UPF packet processing. We conduct an empirical measurement using a commercial 5G core developed from a popular open-source project named free5GC \cite{free_5gc} to understand the LLC usage by these two entities. Surprisingly, we find that these two entities do not use the precious LLC resource properly. As a result, the frequent data eviction from LLC to Dynamic Random Access Memory (DRAM) causes the significant 5G core performance degradation. Three main reasons behind the data eviction are concluded as follows.

\begin{itemize}[leftmargin=*]

\item \textbf{Leaky DMA in DDIO:} The DDIO allows a 5G core to directly write/read packets from NIC to the LLC. Unfortunately, we find that even though some packets are directly sent to the LLC from the NIC, these packets are eventually moved back to DRAM because the CPU does not fetch these packets in time.
Later, when the CPU wants to process these packets, it needs to reload them from DRAM to LLC. \wm{This is called the leaky DMA problem in LLC \cite{resq}}. Our empirical measurements show that the leaky DMA can cause a 5G core throughput degradation by \be{about 20\%}. 

\item \textbf{Hot/cold mbuf in RX buffer:} A 5G core uses RX buffer, which is composed of a number of mbufs, to store packets from/to the NIC. In our 5G core as well as other commercial 5G core \cite{zte_5gc}, the size of the RX buffer is often set with a large value, \eg 262140. Large-size RX buffer is used to reduce packet loss rate, especially when burst traffic shows up. Since the RX buffer has a FIFO ring structure, the \be{tail} of the buffer is always hot because it is recently used due to the FIFO nature. In contrast, the \be{head} of the RX buffer could become cold because it has to wait for a long time before its usage, especially when the size of the RX buffer is large. When the \be{head} becomes cold, its mbuf is evicted from LLC into DRAM \cite{llc-replace-policy}, \be{which can cause about 10\% throughput degradation}.

\item \textbf{LLC cache contention between DDIO and CPU:} We also observe LLC cache contention problem because the LLC is shared across the two entities: DDIO and packet processing. And the competition between these two entities for the LLC resource is directly related to the characteristics of the data flow, \eg packet size. Our empirical measurements show that such cache contention can cause \be{about 10\%$\sim$15\%} throughput degradation when the two entities compete for the same LLC resource.

\end{itemize}

In this paper, we present the design, implementation, and evaluation of \sysname, a system that optimizes cache usage for improving the performance of a 5G core. \sysname identifies three key factors that affect cache miss and hit rate: number of RX descriptors, size of RX buffer, and size of DDIO cache. \sysname reveals how each of the three factors causes severe DMA leakage, hot/cold mbuf issues, and cache contention.

Knowing the root cause of incorrect LLC usages, \sysname designs three modules to optimize cache usage and improve 5G core performance. In the first module, \sysname profiles the LLC status of UPF application to detect incorrect LLC usage problems.
In the second module, to address leaky DMA and hot/cold mbuf problems, 
 \sysname conducts an offline search for the optimal configurations of UPF.
In the thrid module, \sysname dynamically adjusts the size of the DDIO cache and the core cache to minimize the cache contention between them when handling traffic with different sizes of packets.



We deploy and evaluate \sysname with a commercial-version free5GC. We also compare with cache optimization system designed for data centers, such as ResQ \cite{resq}. We have the following conclusion.

\squishlist
\item \sysname is the first system that reveals the root cause of inefficient cache usage and cache contention in 5G UPF. This analysis opens up opportunities to improve 5G UPF performance from the perspective of cache optimization.

\item Compared to the default configuration of today's commercial 5G core, \sysname increases the throughput of a 5G UPF from \wm{54.81Gbps to 76.41Gbps, 39.41\% improvement}. 
Remember, such throughput improvement does not require changing any source code of the 5G core software. 

\item We also compare with cache optimization systems designed for data centers, such as ResQ. The throughput achieved by adopting \sysname is \wm{76.41Gbps, 29.55\% higher than the 58.98Gbps of ResQ}. Moreover, \sysname's advantage retains when the 5G core serves  traffic with different sizes of packets.

\squishend

We will open source \sysname and test data set. This work conforms to the IRB policies (if any) of our institution.

\section{Background}
\label{sec:background}

{
\setlength{\belowcaptionskip}{-10pt}

\begin{figure}[t]
  \centering
  \includegraphics[width=0.9\linewidth]{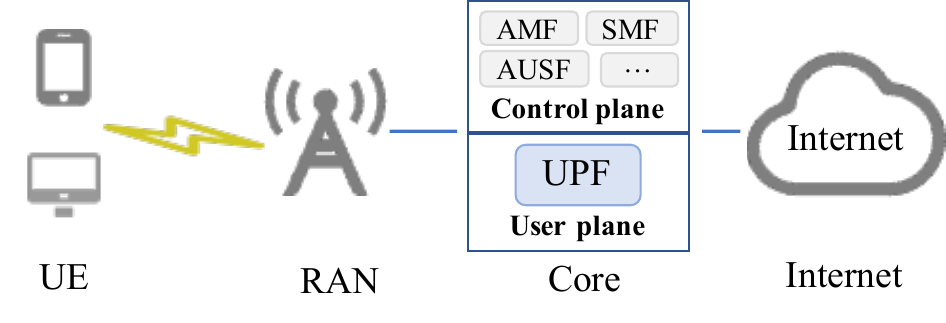}
  \caption{5G architecture.}
  \label{fig:5gc_arch}

\end{figure}
}

\subsection{5G Background}
\label{sec:5g_background}
The architecture of a 5G network consists of two main components, the Radio Access Network (RAN) and the Core network \cite{5g-background1,5g-background2,5g-background3,spacecore}. Figure~\ref{fig:5gc_arch} illustrates a 3GPP-compliant 5G architecture, where User Equipments (UEs) connect to the 5G core through the RAN. The 5G core is responsible for routing bi-directional packets between the RAN and the Internet \cite{5gc-book}. The main focus of this paper is to analyze the performance bottleneck of the 5G core, which consists of two main components: the control plane and the user plane.

\parabf{Control plane:} The control plane of 5G core utilizes a service-based architecture, which defines network functions based on services such as AMF, SMF, \etc \cite{5g-background4,5gc_up,l25gc,localization_core}. This architecture allows efficient communication, scaling, and agility by decoupling and defining microservices.

\parabf{User plane:} The User Plane Function (UPF) is a crucial component in the user plane, responsible for receiving packets and \wm{matching packets} with the Packet Detection Rules (PDRs) provided by the SMF in the control plane \cite{29.244,23.502}. A UPF consumes lots of resources, including LLC cache, because it determines how a data flow is processed. \wm{Note that 5G core vendors \cite{l25gc, samsung_5gc, zte_5gc} usually adopt kernel-bypass functions, such as DPDK \cite{dpdk}, to improve UPF performance.}

{
\setlength{\belowcaptionskip}{-5pt}
\begin{figure}[t]
  \centering
  \includegraphics[width=0.9\linewidth]{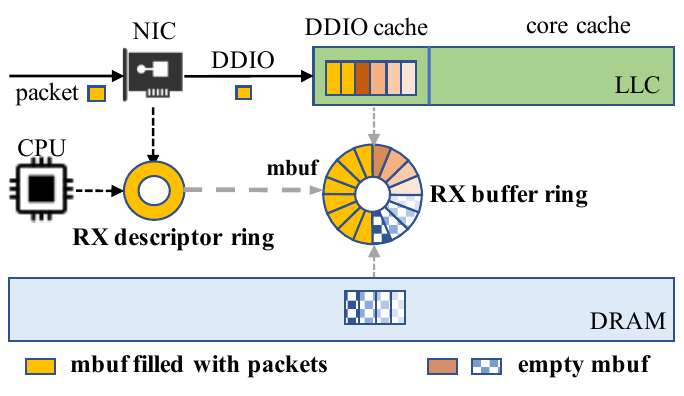}
  \caption{Each RX descriptor points to a memory buffer (mbuf) in the RX buffer. When receiving a packet, DDIO can use the descriptor to directly load the packet into LLC.}
  \label{fig:ring_usage}

\end{figure}
}


{
\begin{figure*}[t]
    \centering
    \subfigure[Packet reception when mbuf stays inside LLC]{
    \begin{minipage}[b]{2.2in}
    \includegraphics[width=2.2in]{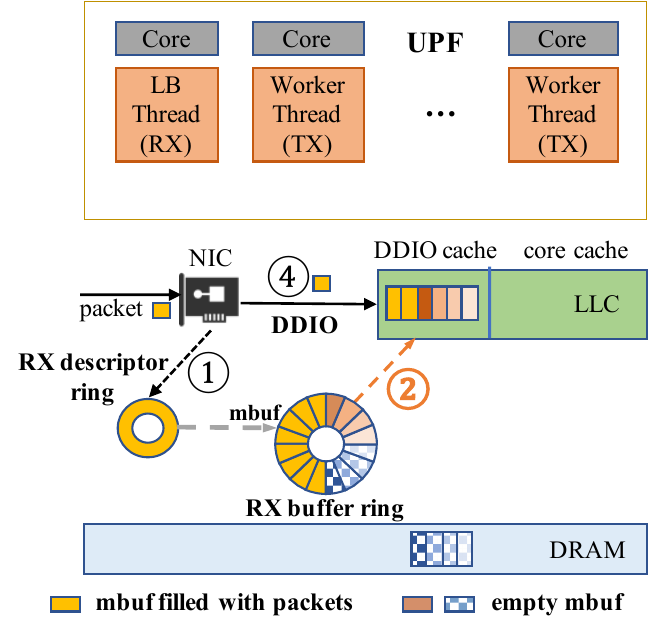}\vspace{-0.2in}
    \label{fig:upf_cache_usage1}
    \end{minipage}
    }
    \subfigure[Packet reception when mbuf stays inside DRAM]{
    \begin{minipage}[b]{2.2in}
    \includegraphics[width=2.2in]{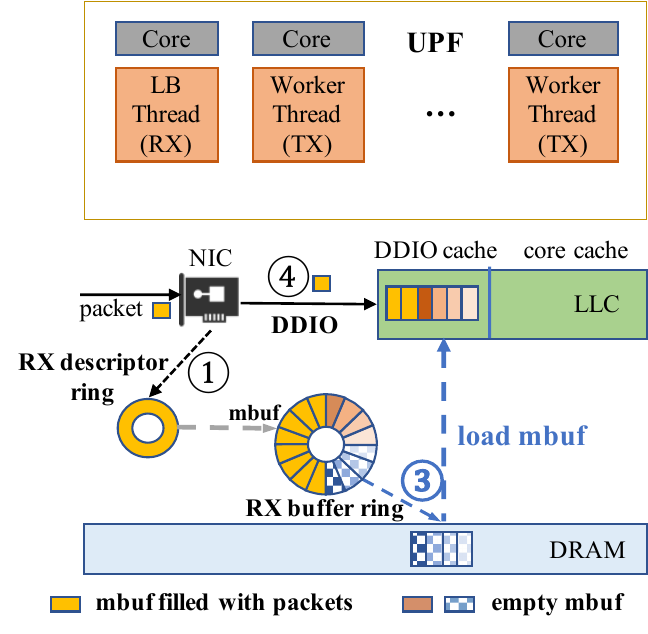}\vspace{-0.2in}
    \label{fig:upf_cache_usage2}
    \end{minipage}
    }
    \subfigure[Packet loading, processing, and transmission]{
    \begin{minipage}[b]{2.2in}
    \includegraphics[width=2.2in]{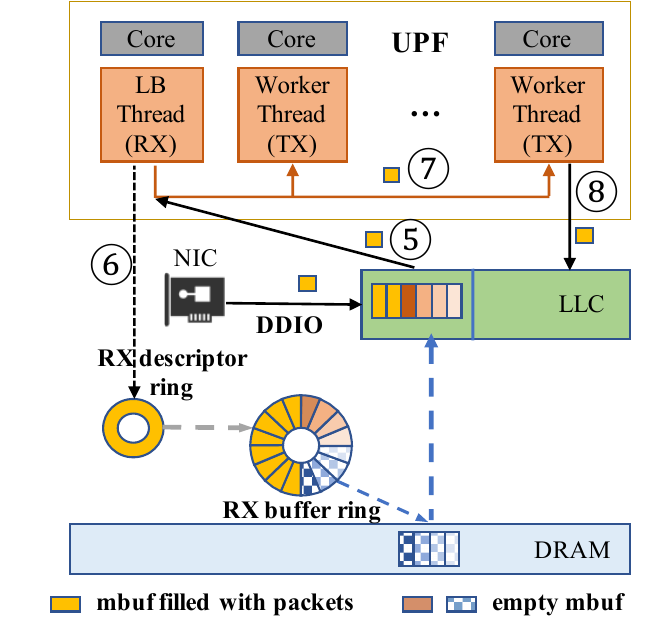}\vspace{-0.2in}
    \label{fig:upf_cache_usage3}
    \end{minipage}
    }
    \caption{Packet reception, loading, processing, and transmission in the 5G UPF.}
    \label{fig:upf_cache_usage}
\end{figure*}
}

\subsection{DDIO, RX buffer and descriptor}
Figure~\ref{fig:ring_usage} shows three key modules that impact the LLC usage in a 5G core: DDIO, RX buffer ring, and RX descriptor ring.

\parabf{DDIO:} DDIO, which stands for Data Direct I/O \cite{ddio}, is a technology developed by Intel that implements Direct Cache Access \be{(DCA)} technology \cite{dca1,dca2,dca3,dca4}. It has been widely adopted in server architectures to enhance network performance \cite{ddio-application1,ddio-application2,ddio-application3}. \wm{As Figure~\ref{fig:ring_usage} shows}, DDIO allows data to be moved directly between the peripherals and the LLC, bypassing \wm{DRAM} and thus reducing latency and the overhead associated with data transfer. Moreover, DDIO improves cache hit rate, enhances efficiency and throughput of packet processing, and effectively addresses the challenges that arise from large-volume data traffic in 5G systems.

\parabf{RX buffer ring:} A key data structure in the 5G core is the RX buffer ring which stores the packets from/to the NIC. To get high performance, a 5G core usually uses a number of \hn{pre-allocated} memory buffers (mbufs) to build the RX buffer ring. The RX buffer is a ring-based structure that utilizes mbufs in a first-in-first-out (FIFO) manner. Each mbuf contains a space to store a packet, a pointer to the next mbuf, and some other information. \hn{As Figure~\ref{fig:ring_usage} shows, the mbufs are either in LLC or DRAM \wm{based on} the cache replacement policy of CPU.}

\parabf{RX descriptor ring:} 
The RX descriptor ring is a critical data structure that enables data exchange between the NIC and CPU core \cite{descriptor}. It comprises a set of descriptors, such as 4096 descriptors in an anonymous vendor's 5G core. Each descriptor includes a pointer to a mbuf, the length of the corresponding packet, and additional information. The mbuf pointer within the descriptor informs the NIC and CPU core where to write/read the packet. For instance, when a data packet arrives, the NIC reads \wm{the descriptor at the head} of the RX descriptor ring and then \hn{writes} the packet \wm{into the mbuf pointed by this descriptor}.


\parabf{LLC in UPF:}
The LLC consists of two distinct parts: the DDIO cache, which is \hn{mainly} responsible for \be{storing packets from/to the NIC}, and the core cache, which stores data used for UPF processing. 

\subsection{Cache usage in 5G core UPF}
\label{sec:cache_usage_upf}
\parabf{How UPF works with LLC?}
5G core vendors like ZTE \cite{zte_5gc} and Samsung \cite{zte_5gc} usually adopt a \textit{pipeline} architecture to implement their 5G core software. \wm{In this architecture, packet reception (RX) and transmission (TX) are implemented in different threads and run on different CPU cores as shown in Figure~\ref{fig:upf_cache_usage}.} \wm{RX (\ie LB thread) handles packet reception and load balancing, while TX (\ie worker thread) is responsible for packet processing and transmission. There are two reasons for choosing pipeline architecture. First, the pipeline separates I/O-\hn{intensive} and CPU-\hn{intensive} functions to achieve high performance. Second, 5G core vendors need to support different enhancing network functions, such as load balancing \cite{5gc_lb} and traffic accounting.} We use Figure~\ref{fig:upf_cache_usage1}, \ref{fig:upf_cache_usage2} and \ref{fig:upf_cache_usage3} to walk through several examples of 5G core functions \be{from the perspective of UPF}. We use \textcircled {x} to indicate the specific operation involved in the workflow of UPF. 




\squishlist
\item \textbf{Packet reception:} As Figure~\ref{fig:upf_cache_usage1} shows, upon the arrival of a data packet, NIC reads \wm{the descriptor at the head} of the RX descriptor ring \textcircled{1} to determine the corresponding mbuf where the packet should be written into. If the mbuf stays inside of LLC \textcircled{2}, then the packet is directly written into the mbuf in LLC by the NIC using DDIO \textcircled{4}. Then the descriptor's status is updated, and the mbuf is marked as \be{filled with packet}.

In the other case, as Figure~\ref{fig:upf_cache_usage2} shows, if the mbuf stays inside of DRAM \textcircled{3}, DDIO has to allocate a new space in LLC for the mbuf.
\be{After the mbuf is allocated in LLC, \hn{following the} previous process, the packet can be directly written into it \textcircled{4}.}

\item \textbf{Loading packet into CPU thread:} As Figure~\ref{fig:upf_cache_usage3} shows, the load balancing (LB) thread retrieves the descriptor to locate {the mbuf filled with packet} in LLC and read the packet from it \textcircled{5}. Once the LB thread receives the data packet, the descriptor is updated to point to a new empty mbuf \textcircled{6}.

\item \textbf{Packet processing and transmission:} As Figure~\ref{fig:upf_cache_usage3} shows, after loading the data packets, the LB thread assigns them to other worker threads for parallel processing \textcircled{7}. These worker threads execute actions, such as forwarding, dropping, buffering, and accounting based on the PDRs matched with the packet \cite{23.501,33.513}. After processing, the worker threads tell the NIC to send the packet. Since the mbuf is recently used, it stays inside of LLC. Therefore, NIC can use DDIO to directly fetch data in the mbuf and send it out \textcircled{8}.
\squishend

\section{Motivation}
\label{sec:motivation}

\sysname's design is motivated by discovering the incorrect LLC usage in today’s commercial 5G core. We find three types of problems: leaky DMA, hot/cold mbuf, and cache contention.

\subsection{Leaky DMA problem} 
\label{sec:leaky_dma}

{

\begin{figure}[t]
  \centering
  \includegraphics[width=0.85\linewidth]{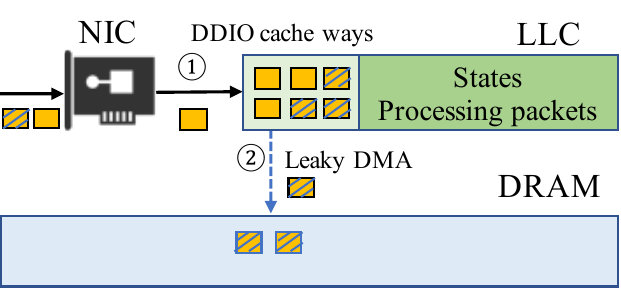}
  \caption{NIC writes packets directly into LLC and then evicts previous packets into DRAM, resulting in DMA leakage.}
  \label{fig:rx_desc}

\end{figure}
}

{
\begin{table}[t]
    \centering
    \normalsize
    \renewcommand{\arraystretch}{1.0}
    \begin{tabular}{lcc}
        \toprule
            & Throughput & Memory bandwidth speed \\ 
            \midrule
            \makecell[l]{Default}       & 54.81Gbps & 14.88GB/s \\ 
            \makecell[l]{ResQ \cite{resq}}       & 58.98Gbps & 13.82GB/s \\ 
            \makecell[l]{$\Delta$}       & 4.17Gbps$\uparrow$ & 1.06GB/s$\downarrow$ \\ 
            \bottomrule
    \end{tabular}
    \caption{UPF performance degradation caused by DMA leakage.}
    \label{tab:leaky_dma_throughput_ddiomiss}
\end{table}
}

The leaky DMA is one example of incorrect LLC usage that causes 5G core performance degradation. Figure~\ref{fig:rx_desc} shows an example of this. In this example, DDIO attempts to transfer packets from NIC into LLC, particularly into the DDIO cache \textcircled {1}. If DDIO finds available space in LLC to store the packet without removing previous packets, the packet is directly written into LLC. This process is called \textit{DDIO write hit}. On the other hand, if DDIO fails to find available space in LLC, 
\hn{this packet will occupy the space in the DDIO cache and evict the previous packet into DRAM \textcircled{2}. This process is called \textit{DDIO write miss}.}

In today's commercial 5G core systems, \hn{LLC in CPU is N-ways associative (\eg 11 cache ways in our testbed)}. The DDIO cache is set to two LLC cache ways \wm{(6MB)}. The size of the DDIO cache is small and usually saturated with packets, especially when the NIC experiences large traffic. As a result, when a new packet arrives, it tends to evict previous packets, resulting in a phenomenon called \textit{leaky DMA} \cite{resq, leakydma2}. This causes packets that were loaded into the DDIO cache to be involuntarily evicted from the LLC to the DRAM by subsequent incoming packets. When a thread needs to process an evicted packet, it must wait for the packet to be reloaded into the LLC, which increases memory bandwidth consumption and leads to performance degradation.



We conduct an empirical measurement with a commercial 5G UPF to understand the impact of leaky DMA. Table~\ref{tab:leaky_dma_throughput_ddiomiss} summarizes the results. Using the default configuration of the commercial 5G UPF, its throughput is 54.81Gbps. The memory bandwidth is 14.88GB/s, which is very large and indicates the presence of DMA leakage because data is moved back and forth between LLC and DRAM. To understand the impact on UPF throughput, we implement ResQ \cite{resq}, a system that addresses the leaky DMA problem. After adopting ResQ, the 5G UPF throughput increases from 54.81Gbps to 58.98Gbps and memory bandwidth speed drops from 14.88GB/s to 13.82GB/s. This result tells us that DMA leakage does exist and can cause UPF performance degradation. 

\subsection{Hot/cold mbuf problem}
\label{sec:cold_mbuf}

{
\setlength{\belowcaptionskip}{-15pt}
\begin{figure}[t]
  \centering
  \includegraphics[width=0.85\linewidth]{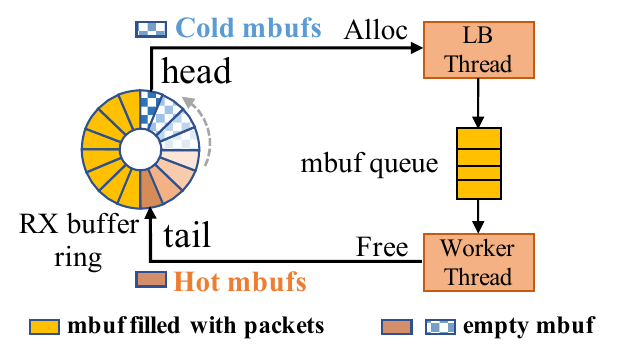}
  \caption{The mbuf becomes cold since not been accessed for a while, resulting in hot/cold mbuf issues.}
  \label{fig:mem_pool_upf}
\end{figure}
}

The hot/cold mbuf problem in a 5G UPF leads to performance degradation \wm{as well}. 
As we discussed in \S\ref{sec:cache_usage_upf}, the LB thread retrieves the RX descriptor and reads the mbuf filled with the packet. Then, as shown in Figure~\ref{fig:mem_pool_upf}, the LB thread will fetch a new empty mbuf from the head of the RX buffer ring to update the RX descriptor. Subsequently, the mbuf filled with the packet received by the LB thread is added to a mbuf queue. 
The worker thread extracts a mbuf from this mbuf queue, processes and transmits it back to the NIC. The worker thread then returns the mbuf to the \wm{tail} of the RX buffer ring.

As depicted in Figure~\ref{fig:mem_pool_upf}, the mbuf located at the \hn{tail} of the RX buffer is considered hot because it was recently used \be{and freed} by the worker thread and therefore resides within the LLC. Conversely, the mbuf located at the \hn{head} of the RX buffer ring is cold since it waits for a while and will be flushed from LLC into DRAM before being utilized by the LB thread. 
The above hot/cold mbuf issues are exacerbated when the size of the RX buffer is greater than the size of the LLC \cite{multithread_dpdk}, a configuration commonly used in commercial 5G UPF systems to reduce packet loss rates, particularly during bursts of traffic. 


{
\begin{table}[t]
    \centering
    \normalsize
    \renewcommand{\arraystretch}{1.0}
    \begin{tabular}{lcc}
        \toprule
        & Throughput & DDIO write miss rate \\ 
        \midrule
        \makecell[l]{Default}       & 65.36Gbps & 96.67\% \\ 
        \makecell[l]{Optimal}       & 71.49Gbps & 67.63\% \\ 
        \makecell[l]{$\Delta$}       & 5.13Gbps$\uparrow$ & 30.04\%$\downarrow$ \\ 
        \bottomrule
    \end{tabular}
    \caption{UPF performance degradation caused by hot/cold mbuf issues.}
    \label{tab:mbuf_throughput_ddiomiss}
\end{table}
}

Table~\ref{tab:mbuf_throughput_ddiomiss} summarizes the results of an empirical measurement with a commercial 5G UPF. Because the mbuf at the \wm{head} of the RX buffer ring is cold and needs to be reloaded into LLC from DRAM, the throughput achieved by the commercial 5G UPF \wm{is 65.36Gbps and the rate of reloading the mbuf (\ie DDIO write miss rate) at the \wm{tail} of the RX buffer ring is 96.67\%. If the size of the RX buffer is chosen properly (\ie optimal in Table~\ref{tab:mbuf_throughput_ddiomiss}), we can reduce the mbuf reloading rate to 67.63\% and increase the throughput to 71.49Gbps.}

\subsection{Cache contention problem} 
\label{sec:cache_contention}




As we discussed in \S\ref{sec:5g_background}, the 5G core consists of the control plane and the user plane. The control plane includes several network elements, such as SMF and AMF. The user plane is mostly UPF, where UPF-CP is the control plane of UPF itself.
We measure the LLC occupation of different NFs using a commercial 5G UPF, as shown in Figure~\ref{fig:eva_upf_5gc_nf_llc_occupy}. 
We can observe that the LLC occupation of UPF increases from 1.28MB to 20.45MB as UPF's throughput increases from 5Mbps to 5000Mbps. 
However, other threads (\eg UPF-CP, SMF, and AMF) keep a low and stable LLC occupation (less than 500KB) as the throughput increases. 
Therefore, we can conclude that UPF uses most of LLC. 

The cache contention problem in a 5G UPF also leads to performance degradation. Figure~\ref{fig:ddio_core} shows an example of this problem. Although LLC is divided into DDIO cache and core cache, in state-of-the-art commercial systems, LLC resources are shared among all CPU cores in the same CPU socket. As a result, the CPU core could use DDIO cache resources for packet processing. Similarly, NIC could use core cache for storing ingress packets.



{
\setlength{\abovecaptionskip}{-0pt}
\setlength{\belowcaptionskip}{-10pt}

\begin{figure}[t]
  \centering
  \includegraphics[width=0.95\linewidth]{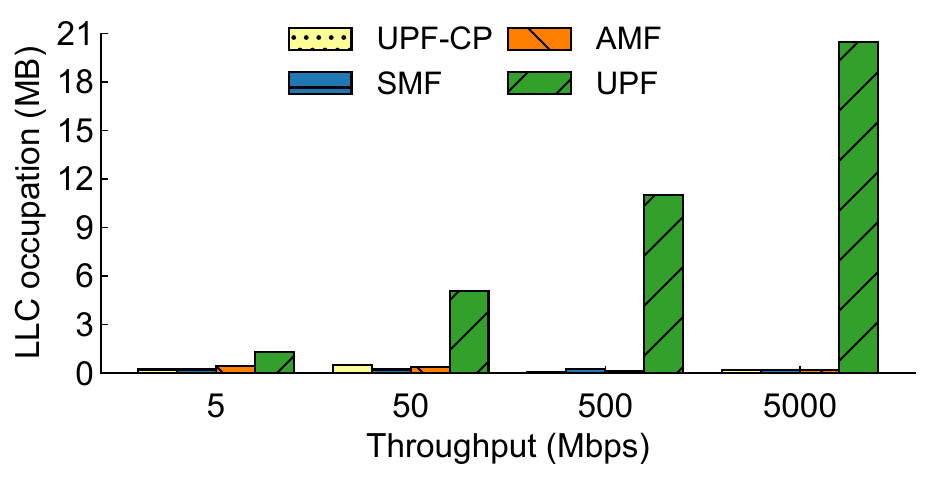}
  \caption{LLC usage by different network elements in a 5G core.}
  \label{fig:eva_upf_5gc_nf_llc_occupy}

\end{figure}
}

{
\setlength{\abovecaptionskip}{-0pt}
\setlength{\belowcaptionskip}{-0pt}

\begin{figure}[t]
  \centering
  \includegraphics[width=0.85\linewidth]{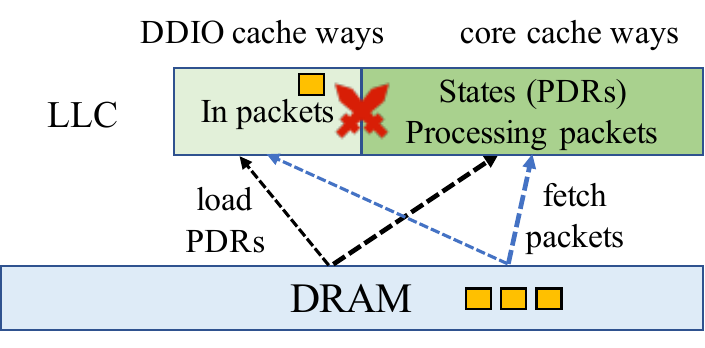}
  \caption{DDIO and CPU core compete for the same LLC resource. This is called cache contention.}
  \label{fig:ddio_core}

\end{figure}
}

{

\begin{table}[t]
    \centering
    \normalsize
    \renewcommand{\arraystretch}{1.0}
    \begin{tabular}{lccc}
        \toprule
        & Throughput & \makecell[l]{RX core \\ LLC miss rate} & \makecell[l]{TX core \\ LLC miss rate} \\ 
        \midrule
        \makecell[l]{Default}       & 54.81Gbps & 63.62\% & 30.71\% \\ 
        \makecell[l]{Isolation}       & 61.72Gbps & 46.72\% & 26.05\% \\ 
        \makecell[l]{$\Delta$}       & 6.91Gbps$\uparrow$ & 16.90\%$\downarrow$ & 4.66\%$\downarrow$ \\ 
        \bottomrule
    \end{tabular}
    \caption{UPF performance degradation caused by cache contention.}
    \label{tab:contention_throughput_ddiomiss}
\end{table}
}



\be{Table~\ref{tab:contention_throughput_ddiomiss} summarizes the results of our empirical measurement. Because of the contention between DDIO cache and core cache, the throughput achieved by the commercial 5G UPF is 54.81Gbps. The LLC miss rates of the reception thread (RX) and transmission thread (TX) are 63.62\% and 30.71\%, indicating IO and CPU core using DDIO cache and core cache, respectively. In this case, we configure DDIO cache and core cache in isolation mode, which can reduce cache contention. We can observe that the throughput is increased to 61.72Gbps while RX and TX LLC miss rates are reduced to 46.72\% and 26.05\%, respectively.}

\section{\sysname Overview}
\label{sec:overview}

{
\setlength{\abovecaptionskip}{+2pt}
\setlength{\belowcaptionskip}{-8pt}

\begin{figure}[t]
  \centering
  \includegraphics[width=0.9\linewidth]{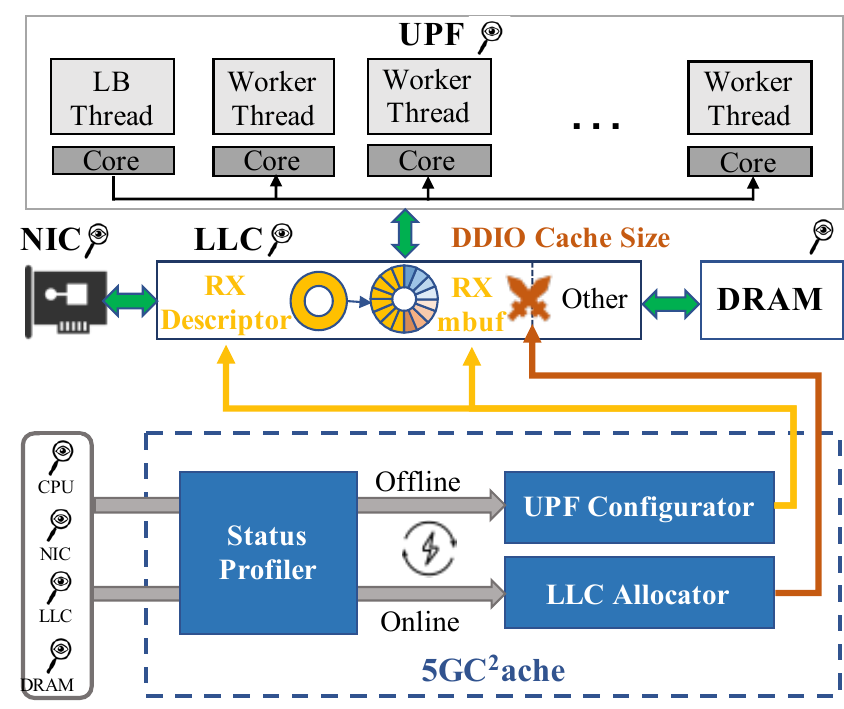}
  \caption{\sysname system overview.}
  \label{fig:overview}

\end{figure}
}

Figure~\ref{fig:overview} shows an overview of \sysname which has three modules.

{

\begin{figure*}[t]
    \centering
    \subfigure[Memory bandwidth speed]{
    \begin{minipage}[b]{2.2in}
    \includegraphics[width=2.2in]{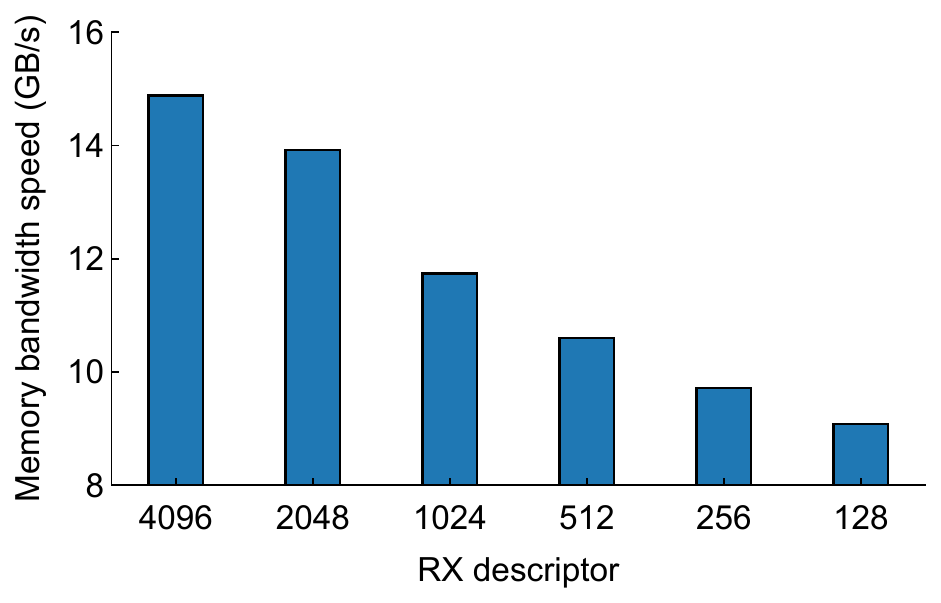}\vspace{-0.2in}
    \label{fig:eva_upf_rx_desc_read_bw_throughput_moti3}
    \end{minipage}
    }
    \subfigure[DMA leakage ratio]{
    \begin{minipage}[b]{2.2in}
    \includegraphics[width=2.2in]{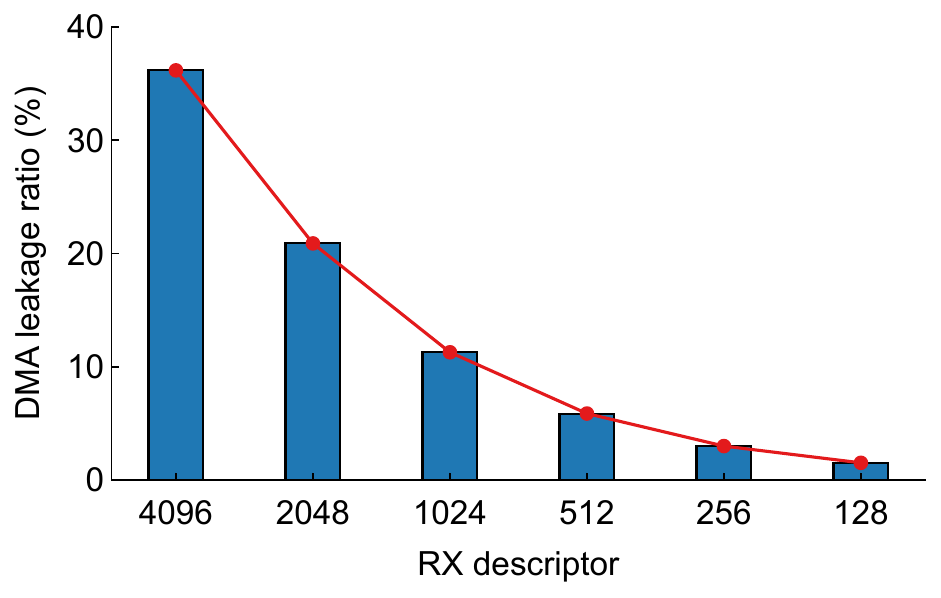}\vspace{-0.2in}
    \label{fig:eva_hash_collision_new}
    \end{minipage}
    }
    \subfigure[Packet loss rate]{
    \begin{minipage}[b]{2.2in}
    \includegraphics[width=2.2in]{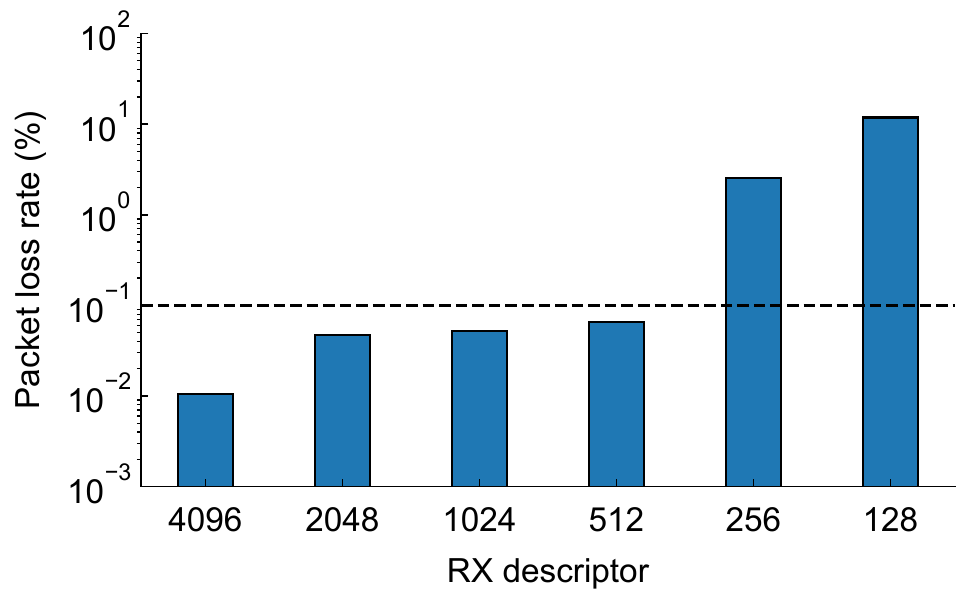}\vspace{-0.2in}
    \label{fig:eva_upf_rx_desc_pkt_loss_log_moti}
    \end{minipage}
    }
    \caption{UPF performance with different numbers of RX descriptors.}
    \label{fig:leaky_dma}
\end{figure*}
}

\squishlist

\item \textbf{Status Profiler:}
The status profiler module keeps tracking on many factors that affect the performance of UPF data plane, including the workloads on UPF cores, the traffic on the NIC, the LLC stats of UPF threads and the DRAM bandwidth consumption.
\item \textbf{UPF Configurator:}
The statistical data gathered by the status profiler can inform the configurations of UPF, including the size of RX descriptors and RX buffer ring, which are the main causes of leaky DMA and hot/cold mbuf problems.
Since the static configurations of UPF can not be modified during runtime, the UPF configurator conducts an offline search for the optimal settings.
\item \textbf{LLC Allocator:}
To deal with the cache contention problem between the DDIO and UPF core cache ways under varying traffic patterns, the LLC allocator leverages the statistical data gathered by the status profiler and dynamically adjusts the LLC allocations during runtime.

\squishend

\section{\sysname Design}
\label{sec:design}

\subsection{UPF Configurator}
\parabf{Alleviating DMA leakage} 
\label{sec:rx_desc_sec}
The size of the RX descriptor is a major factor that impacts the severeness of DMA leakage. Figure~\ref{fig:eva_upf_rx_desc_read_bw_throughput_moti3} shows our empirical measurement with a commercial 5G UPF. We can observe that when the RX descriptor size is 4096, the memory bandwidth speed is 14.88GB/s. When we reduce the size to 128, the memory bandwidth speed reduces to 9.09GB/s, smaller than that of 4096 descriptors. 

We derive a theoretical model to understand the probability of DMA leakage. In this theoretical model, we assume that all the ingress packets are uniformly and randomly loaded into LLC. The expectation of DMA leakage $\mathbb{E}(\boldsymbol{Y})$ can be calculated using the following equation where $N$ is the size of data and $M$ is the size of LLC cache. Details of the derivation can be found in Appendix~\ref{apx:hash_collision}.
\begin{align}
\label{eq:hash_collision}
    \mathbb{E}(\boldsymbol{Y})=N-M+M\left (1-\frac{1}{M}\right )^{N}
\end{align}

Figure~\ref{fig:eva_hash_collision_new} shows that the theoretical DMA leakage rate is 36.16\% when the number of RX descriptors is 4096. When we reduce the number to 128, the theoretical DMA leakage rate drops to 1.51\%, much smaller than the case of 4096 descriptors. 

The reason behind this is that a larger number of RX descriptors leads to more packets being loaded into the LLC, while a smaller number of RX descriptors results in the opposite. When a large number of packets are loaded into the LLC, the probability of DMA leakage increases because incoming packets have a higher probability of evicting previous packets from the LLC. Thus, the rate of DMA leakage increases. 

{
\begin{figure*}[t]
    \centering
    \subfigure[UPF DDIO write miss rate]{
    \begin{minipage}[b]{2.2in}
    \includegraphics[width=2.2in]{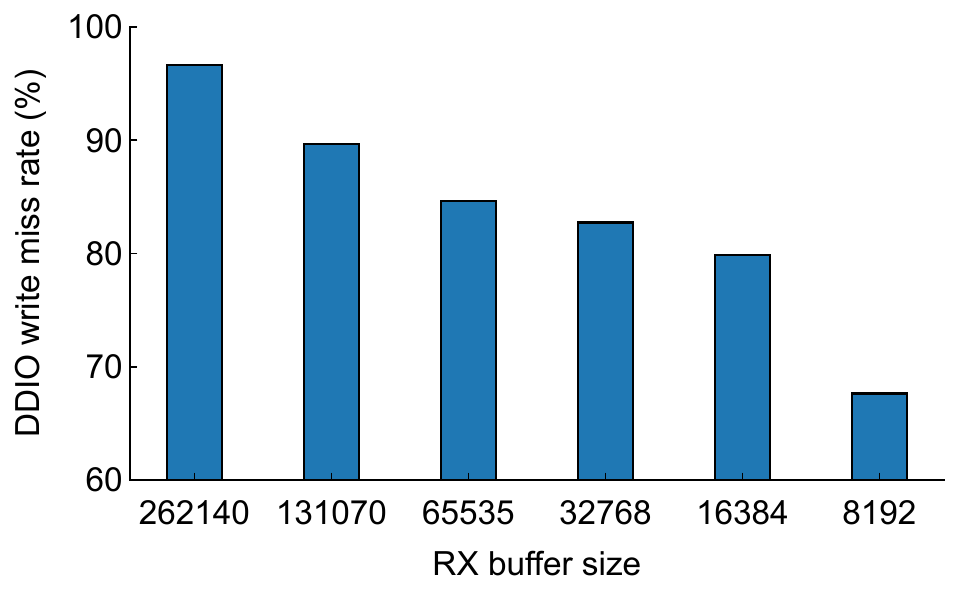}\vspace{-0.2in}
    \label{fig:eva_upf_mem_pool_ddio_wrhit_rate_moti3}
    \end{minipage}
    }
    \subfigure[UPF throughput]{
    \begin{minipage}[b]{2.2in}
    \includegraphics[width=2.2in]{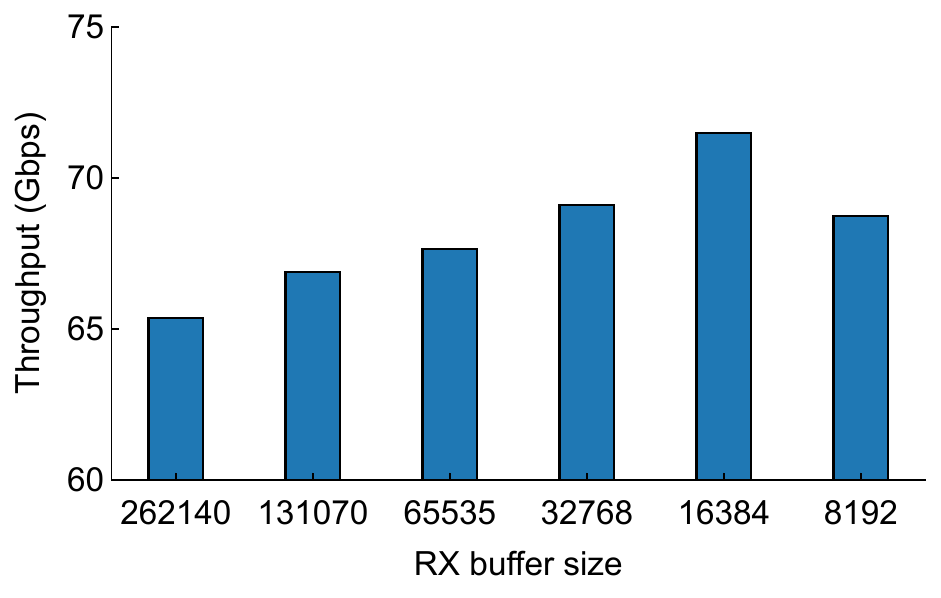}\vspace{-0.2in}
    \label{fig:eva_upf_mem_pool_throughput_moti3}
    \end{minipage}
    }
    \subfigure[Mbuf shortage in UPF]{
    \begin{minipage}[b]{2.2in}
    \includegraphics[width=2.2in]{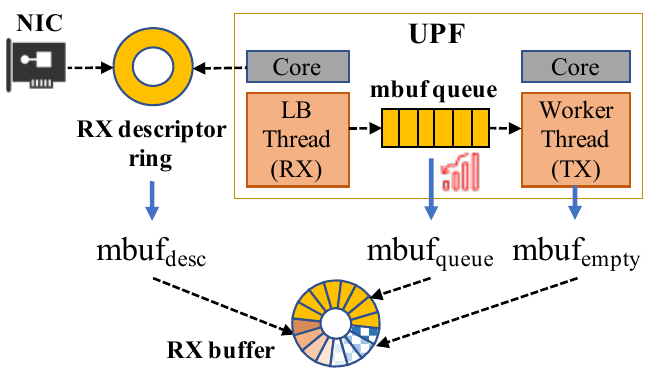}\vspace{-0.1in}
    \label{fig:rx_buffer_throughput_decrease}
    \end{minipage}
    }
    \caption{UPF performance with different RX buffer sizes.}
    \label{fig:hot_mbuf}
\end{figure*}
}

Both Figure~\ref{fig:eva_upf_rx_desc_read_bw_throughput_moti3} and \ref{fig:eva_hash_collision_new} suggest that a smaller number of RX descriptors is better for eliminating DMA leakage. However, when choosing a small number of RX descriptors, we observe a higher packet loss rate. Figure~\ref{fig:eva_upf_rx_desc_pkt_loss_log_moti} shows the packet loss rate measured with different numbers of RX descriptors. The packet loss rate is 0.01\% when we have 4096 descriptors and 11.84\% when 128 descriptors are used. 

When the number of RX descriptors is small, the NIC does not have sufficient  RX descriptors to write the incoming packets into the RX buffer. As a result, the NIC is forced to drop packets at its ingress port and the packet loss rate increases \cite{packetloss1,packetloss2}. When the number of RX descriptors is large, the packet loss rate is only 0.01\%$\sim$0.07\% because the NIC have sufficient \wm{RX descriptors} to write the incoming packets into the RX buffer.

Considering both DMA leakage and packet loss rate, we need to select the optimum number of RX descriptors to balance the two factors. The UPF configurator selects the size of RX descriptors in a way such that the packet loss rate is not high and DMA leakage is not severe. As a result, the throughput of the overall system is optimized.

\parabf{Resolving hot/cold mbuf issues}
\label{sec:rx_buffer_sec}
The size of the RX buffer plays a key role in the hot/cold mbuf issues. Figure~\ref{fig:eva_upf_mem_pool_ddio_wrhit_rate_moti3} shows our empirical measurement using a commercial 5G UPF. We can observe that when the RX buffer size is 262140, the DDIO miss rate caused by the hot/cold mbuf issues is 96.67\%. When we reduce the RX buffer size to 8192, the DDIO miss rate reduces to 67.63\%, smaller than that of 262140 RX buffer size. When the RX buffer is large, the packet reception thread has a higher probability of getting a cold mbuf because this thread obtains mbufs from the head of the RX buffer, which is likely to be cold due to the large RX buffer size. A smaller RX buffer size leads to the opposite conclusion. 

Figure~\ref{fig:eva_upf_mem_pool_ddio_wrhit_rate_moti3} indicates that a smaller RX buffer size is better for alleviating hot/cold mbuf. However, when we choose a small RX buffer size, we observe a lower throughput. Figure~\ref{fig:eva_upf_mem_pool_throughput_moti3} shows the UPF's throughput measured with different sizes of RX buffer. When the RX buffer size is 8192, the throughput is 68.74Gbps, which is less than 71.49Gbps when the RX buffer size is 16384. 

The root cause of throughput degradation is the mbuf shortage issue. To understand this problem, we use Figure~\ref{fig:rx_buffer_throughput_decrease} to show how UPF uses mbuf in the RX buffer. The RX buffer contains three types of mbufs: mbufs loaded with NIC packets via descriptors mbuf$_{desc}$, mbufs in the mbuf queue between the packet reception thread (RX) and the transmission thread (TX) mbuf$_{queue}$, and unused mbufs mbuf$_{empty}$. Therefore, the size of RX buffer $RX_b$ can be derived using the following equation.
\begin{align}
\label{eq:rx_buffer1}
   RX_b= mbuf_{desc} + mbuf_{queue} + mbuf_{empty}
\end{align}

When the RX buffer size $RX_b$ decreases, both mbuf$_{queue}$ and mbuf$_{empty}$ also decrease. When mbuf$_{queue}$ gets smaller, the queue between the transmission and reception threads becomes shorter. As a result, the UPF could experience cases where the transmission thread depletes all the packets in the queue, but the reception thread does not have a sufficient speed to fill the queue. \be{Consequently}, the overall throughput decreases because the transmission thread does not have data for transmission. This phenomenon becomes severe, especially when the queue size is small.

Therefore, when the RX buffer is large, the UPF experiences a higher DDIO miss rate caused by the hot/cold mbuf issues. When the RX buffer is small, the UPF throughput suffers due to a depleted queue between the packet transmission and reception threads. The UPF configurator conducts an offline profiling to search an optimal RX buffer size.

\subsection{LLC Allocator}
\label{sec:decision_making_sec}




\parabf{Why should we adjust the size of DDIO cache?}
When DDIO cache and core cache are configured in shared mode, we observe severe cache contention between DDIO and CPU core as described in \S~\ref{sec:cache_contention}. \sysname addresses this problem by configuring both the DDIO cache and core cache in isolation mode and adjusting the size of DDIO cache. When both caches work in isolation mode, the contention is no longer severe. However, if we do not dynamically adjust the size of DDIO cache when it stays in isolation mode, the performance of a UPF is bad because it needs to serve different types of traffic.

Figure~\ref{fig:eva_upf_cache_contention_packet_size_cacheways} shows the optimal DDIO cache size for a UPF when the ingress traffic has different sizes of packets. When the size of packets in the traffic is 1500 bytes, UPF should use 5 DDIO cache ways to achieve the optimal throughput. When the size is 256 bytes, UPF should use 3 DDIO cache ways. Therefore, the size of DDIO cache should be dynamically adjusted to obtain optimal performance, especially when the ingress packets have various sizes.

{

\begin{figure}[t]
  \centering
  \includegraphics[width=0.7\linewidth]{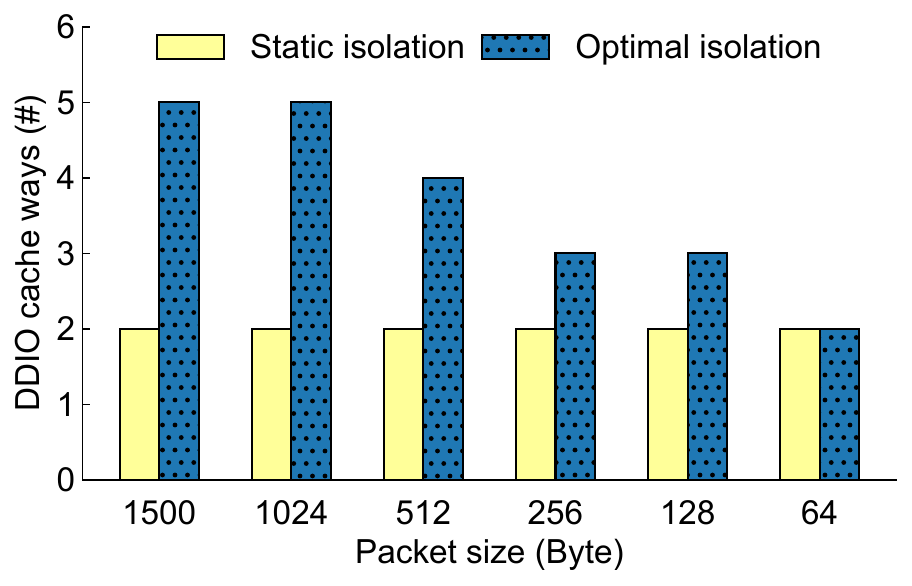}
  \caption{Optimal DDIO cache ways for UPF with different types of traffic.}
  \label{fig:eva_upf_cache_contention_packet_size_cacheways}

\end{figure}
}

\parabf{How to adjust the size of DDIO cache?}
To determine the size of DDIO cache, we need to identify between DDIO and CPU core, which needs more cache. If one of the two needs more cache, \sysname adjusts the DDIO cache size to meet the demand. When both DDIO and CPU core are short of cache, then \sysname keeps the current cache allocation.

To identify whether DDIO cache and core cache run out of capacity, we empirically measure the following two metrics: DDIO write miss and LLC miss. DDIO write miss is calculated by measuring the rate DDIO cannot find the data packet in the DDIO cache. Similarly, the LLC miss rate is calculated by counting the cases where the CPU core cannot find the cache line from LLC when reading data from the cache. These two metrics tell us whether the DDIO cache and the core cache are crowded. 

\sysname uses the state machine in Figure~\ref{fig:state_machine} to dynamically adjust the size of DDIO cache, core cache, and others cache (cache used by control plane and \etc) in its corresponding state. \sysname is aimed at keeping two states (\ie no bottleneck and DDIO-core cache balance) to achieve UPF's high performance.

For every time interval $t_n$, \sysname collects the following three information: \textit{PCIE\_bandwidth}, \textit{DDIO\_write\_miss\_rate} and 
\textit{LLC\_miss\_rate}. Then, \sysname compares the three current data with information collected in the previous time interval $t_{n-1}$. \hn{If these metrics change evidently, the state machine starts working.}
\sysname divides the states of UPF into four types, which are described as follows.

\squishlist

\item \textbf{No bottleneck:} 
This state indicates UPF serves low traffic (corresponds to low PCIe bandwidth) without bottlenecks. As Figure~\ref{fig:state_machine} and Table~\ref{tab:operation_actions} show, this state could transfer to DDIO cache bottleneck state \textcircled{2} (\ie PCIe bandwidth of UPF is higher than the PCIe bandwidth threshold). 

\item \textbf{DDIO-core balance:} 
This is a state where \sysname finds a trade-off between DDIO cache and core cache. As the traffic changes, we can observe that this state could transfer to three potential states, that is, DDIO cache bottleneck \textcircled{4}, no bottleneck \textcircled{5}, and core cache bottleneck \textcircled{6}. 

\item \textbf{DDIO cache bottleneck:} 
This state indicates that UPF faces DDIO cache bottleneck, which means a high DDIO write miss rate. \be{As the traffic changing or cache adjusting}, it could transfer to two potential states: no bottleneck \textcircled{1} and DDIO-core cache balance \textcircled{3}. 

\item \textbf{Core cache bottleneck:} 
This state indicates that UPF faces the core cache bottleneck, that is, the LLC miss rate is high. \be{As the traffic changing or cache adjusting}, this state could transfer to two potential states: DDIO-core cache balance \textcircled{7} and no bottleneck \textcircled{8}. 

\squishend

{

\begin{figure}[t]
  \centering
  \includegraphics[width=0.9\linewidth]{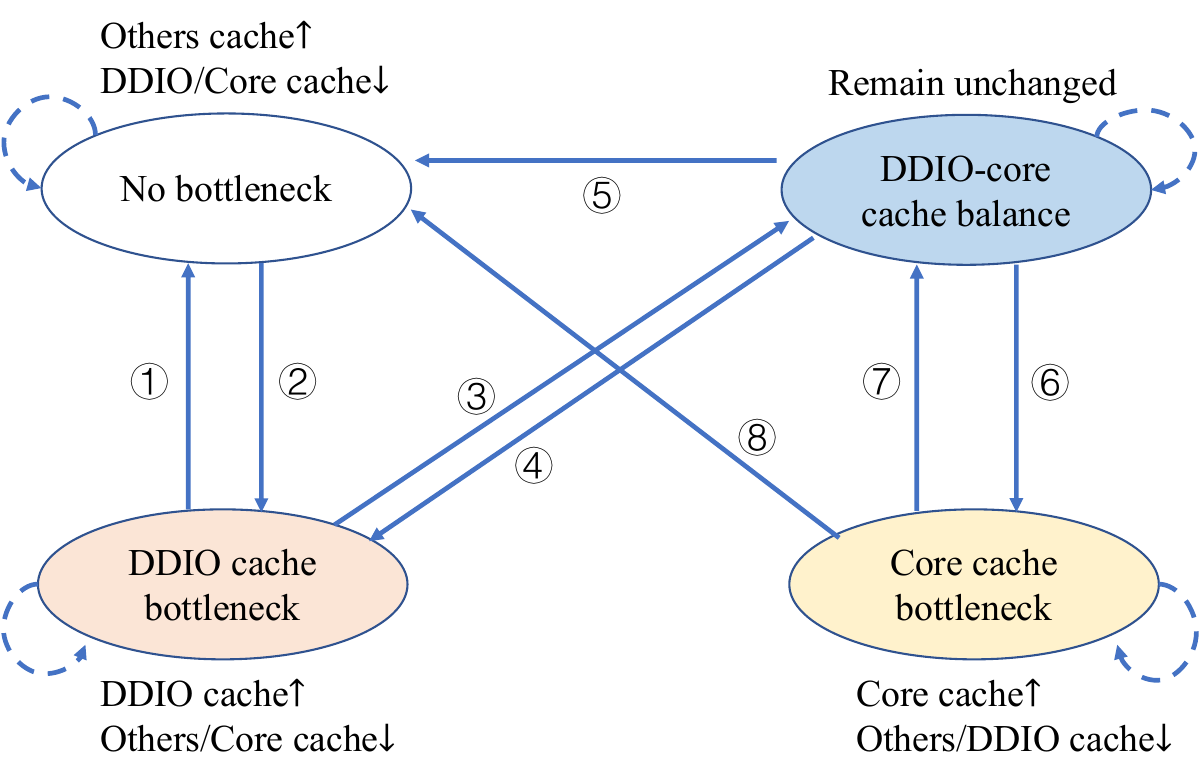}
  \caption{State machine identifies different types of bottleneck to reduce cache contention.}
  \label{fig:state_machine}

\end{figure}
}

{

\begin{table}[t]
    \centering
    \normalsize
    \renewcommand{\arraystretch}{1.0}
    \begin{tabular}{ll}
        \toprule
            State transition & Requirements \\ 
            \midrule
            \makecell[l]{$\textcircled{2}$}       & $pcie\_bandwidth(t) >  PCIE\_BW\_THR $ \\
            \makecell[l]{$\textcircled{1}, \textcircled{5},\textcircled{8}$}       & $pcie\_bandwidth(t) \leq PCIE\_BW\_THR $  \\
            \makecell[l]{$\textcircled{3}$}       & $DDIO\_write\_miss\_rate(t) \to$  \\
            \makecell[l]{$\textcircled{4}$}       & $DDIO\_write\_miss\_rate(t) \uparrow$  \\
            \makecell[l]{$\textcircled{6}$}       & $LLC\_miss\_rate(t) \uparrow$  \\
            \makecell[l]{$\textcircled{7}$}       & $LLC\_miss\_rate(t) \to$ \\
            \bottomrule
    \end{tabular}
    \caption{State transition.}
    \label{tab:operation_actions}
\end{table}
}

\section{Implementation}
\label{sec:implementation}

\sysname's implementation has three components: status profiler, UPF configurator, and LLC allocator, with about 1k, 2k, and 2k lines of code, respectively.

\parabf{Status Profiler:}
The status profiler does two things. 
First, it integrates many cache-related tools, such as PCM \cite{pcm}, Perf \cite{perf}, Intel CAT \cite{cat}, and \etc These tools are used to capture metrics, including PCIe bandwidth, LLC miss/hit, memory bandwidth, and \etc. 
Second, the status profiler implements different threads to call different interfaces of the above tools, so that provides real-time metrics for the next two modules.

\parabf{UPF Configurator:}
In the UPF configurator module, we first code multiple test cases to register multiple UEs and establish corresponding PDU (Packet Data Unit) sessions in the 5G core. 
Second, we implement different traffic trace configurations (\eg traces with different flows, packet sizes and \etc), which are used in Trex\cite{trex} for traffic generation. 
The UPF configurator module feeds the 5G core with different cache-related configurations (\ie number of RX descriptors and size of RX buffer) to different traffic traces. 
In this module, \sysname can choose the optimal number of RX descriptors and size of RX buffer to achieve high performance. 


\parabf{LLC allocator:}
First, we implement a state machine, which reads metrics from the status profiler, analyzes the LLC resource usage in the 5G core system, and transfers state to identify whether DDIO cache and core cache exist bottleneck. 
Second, the state machine re-allocates LLC resources by integrating the Intel CAT tool \cite{cat}, which enables cache ways isolation and allocation. Moreover, to adjust DDIO cache ways, \sysname reads and writes DDIO-related MSRs (Model-specific registers) according to Intel's recommendation \cite{intel_register_doc}. 

\section{Evaluation}
\label{sec:evaluation}
Our evaluation focuses on answering the following four questions:
\squishnumlist
\item What is the performance of a commercial 5G UPF after adopting \sysname (section ~\ref{sec:case_study})?
\item Whether \sysname selects an optimal number of RX descriptors (section~\ref{sec:rx_desc})?
\item Whether \sysname selects an optimal size for RX buffer (section~\ref{sec:mem_pool})?
\item Whether \sysname manages and re-allocates LLC resources to minimize cache contention (section~\ref{sec:ddio_size})?
\squishend

\subsection{Setup and baselines}
\label{sec:setup}

\begin{figure}[t]
  \centering
  \includegraphics[width=1\linewidth]{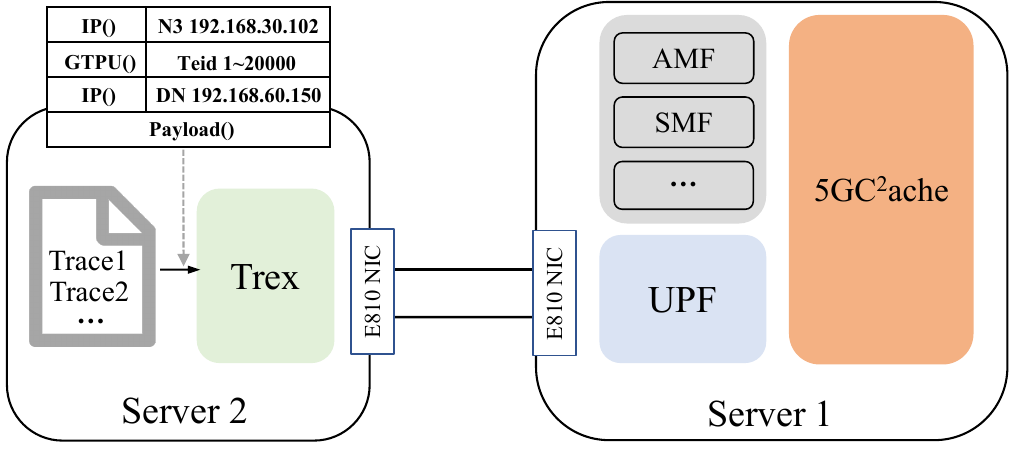}
  \caption{Testbed setup topology.}
  \label{fig:e_setup}
\end{figure}

\parabf{Testbed:} 
Figure~\ref{fig:e_setup} shows the topology of the testbed used in our evaluation. It has two 24-core servers (Intel(R) Xeon(R) Platinum 8163 CPU @ 2.50GHz) running CentOS 7.9. The LLC cache size of a server is 33MB, composed of 11 cache ways. Each server is equipped with an Intel E810 100G Dual Port NIC. We deploy \sysname and a commercial-version 5G core (including AMF, SMF, AUSF, UDM, PCF, UPF-CP, UPF, and \etc) developed from free5GC \cite{free_5gc} on server1. The UPF uses one core for RX and six cores for TX.

The traffic generator is deployed on server2, which produces uplink/downlink traffic. This experiment uses Cisco TRex \cite{trex} to generate traffic. TRex is an open-source software traffic generator capable of generating 100Gbps traffic. 

{
\setlength{\abovecaptionskip}{3pt}

\begin{figure}[t]
  \centering
  \includegraphics[width=1\linewidth]{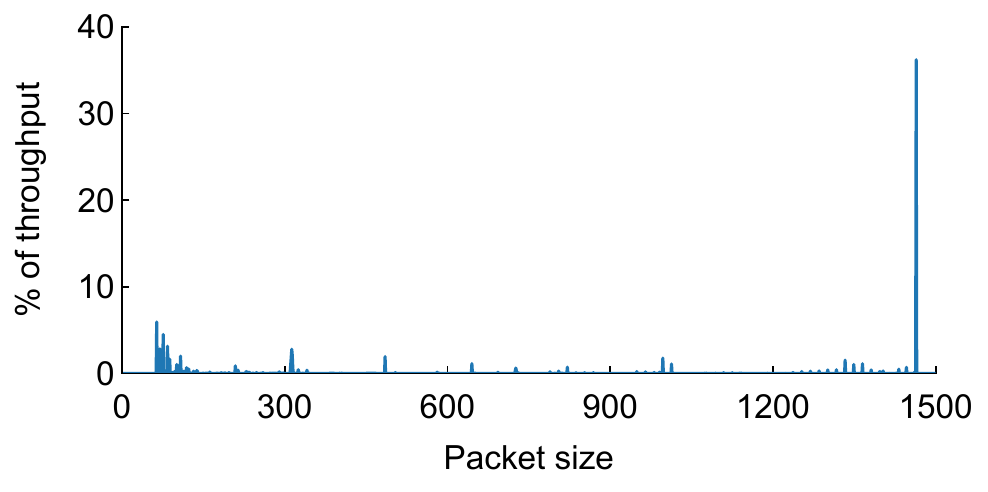}
  \caption{The ratio of different packet sizes to overall throughput in a real trace with 31M packets.}
  \label{fig:eva_cciot_packet_size}

\end{figure}
}

{
\setlength{\abovecaptionskip}{3pt}
\begin{figure*}[t]
    \centering
    \subfigure[UPF throughput]{
    \begin{minipage}[b]{2.2in}
    \includegraphics[width=2.2in]{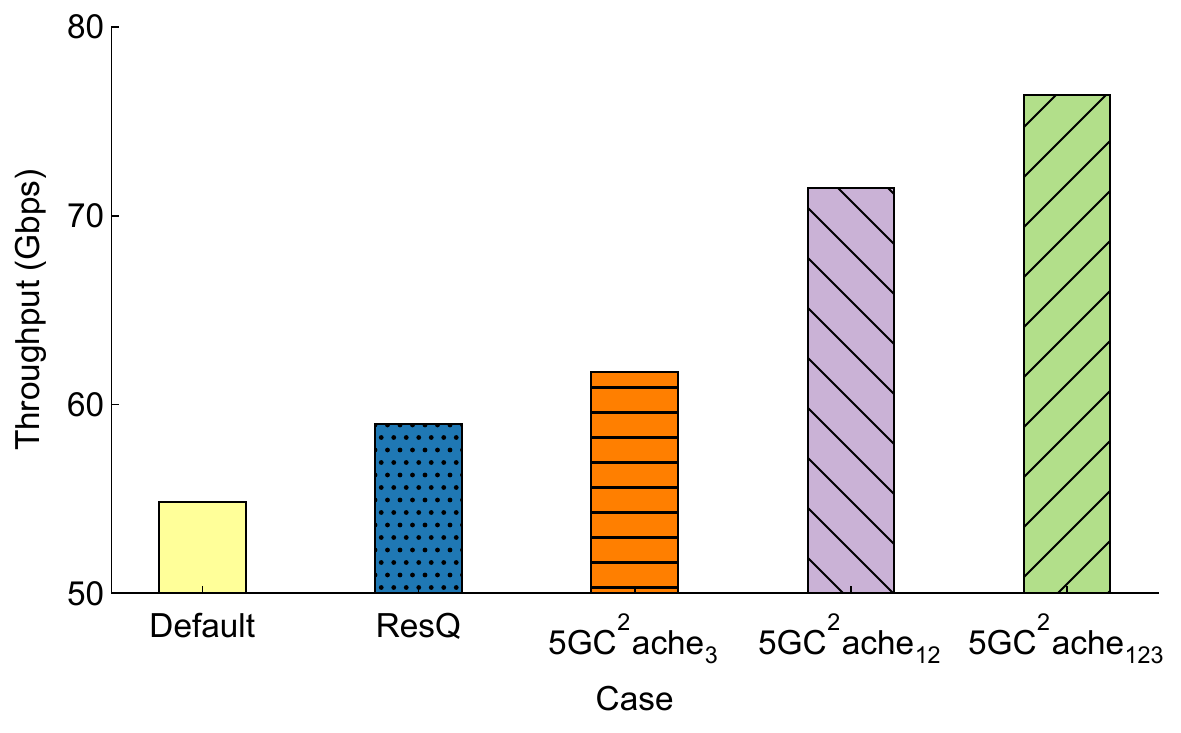}\vspace{-0.2in}
    \label{fig:eva_upf_case_study_throughput}
    \end{minipage}
    }
    \subfigure[UPF RX core LLC miss rate]{
    \begin{minipage}[b]{2.2in}
    \includegraphics[width=2.2in]{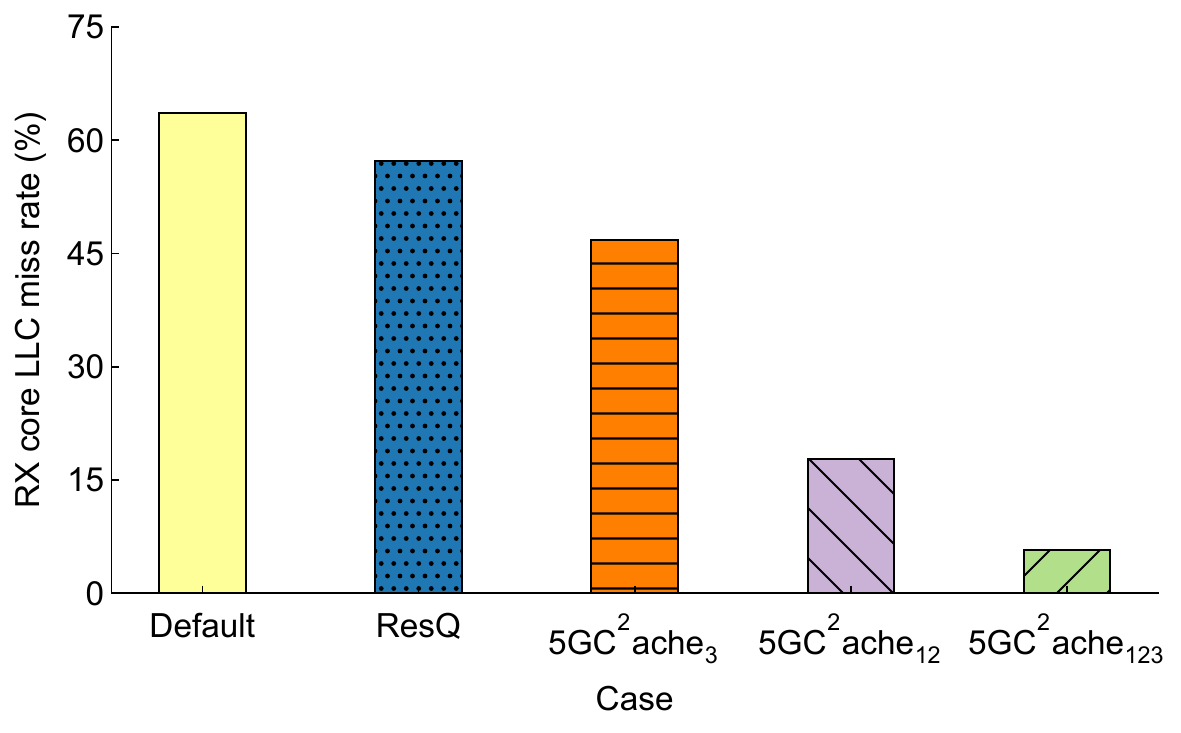}\vspace{-0.2in}
    \label{fig:eva_upf_case_study_rx_llc_miss}
    \end{minipage}
    }
    \subfigure[UPF TX core LLC miss rate]{
    \begin{minipage}[b]{2.2in}
    \includegraphics[width=2.2in]{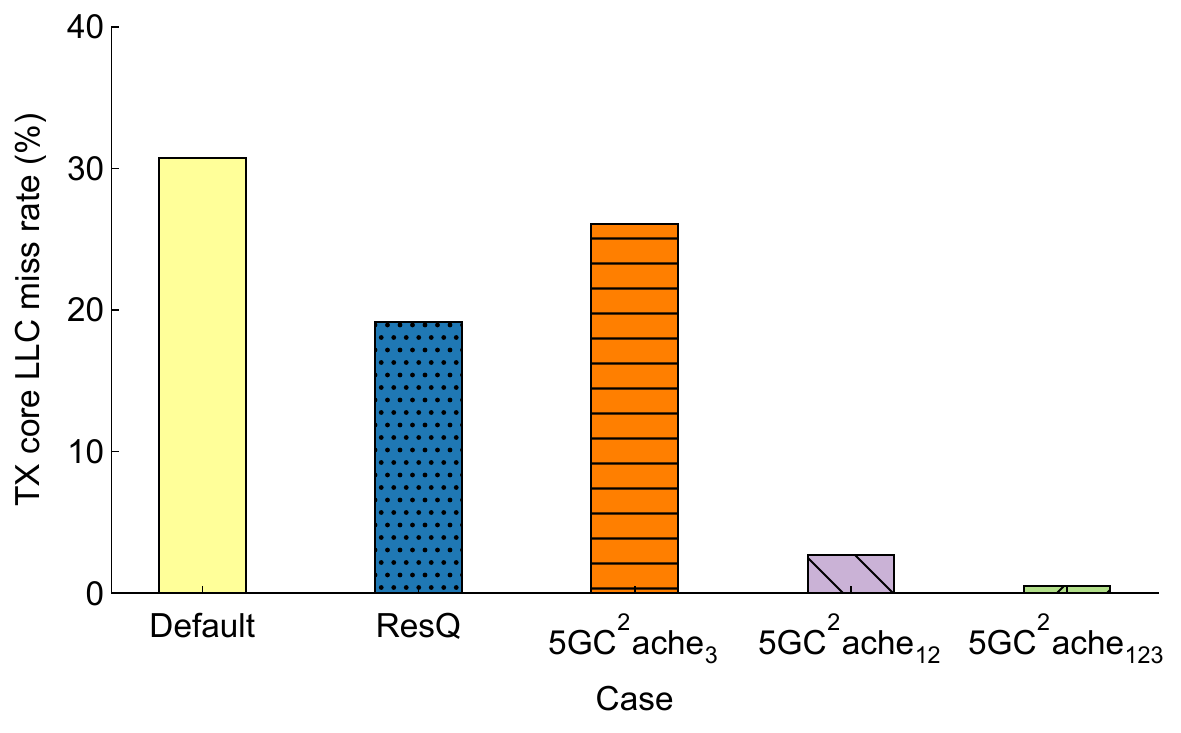}\vspace{-0.2in}
    \label{fig:eva_upf_case_study_tx_llc_miss}
    \end{minipage}
    }
    \caption{UPF performance under different solutions.}
    \label{fig:eva_case_study}
\end{figure*}
}

{

\begin{figure}[t]
  \centering
  \includegraphics[width=0.9\linewidth]{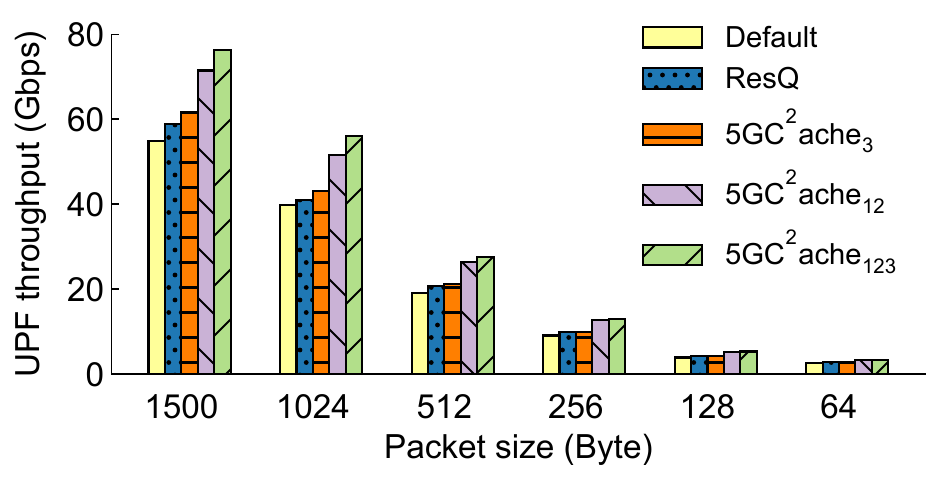}
  \caption{UPF throughput when serving traffic with different sizes of packets.}
  \label{fig:eva_upf_case_study_diff_pkt_size}

\end{figure}
}

\parabf{UE and RAN simulation:}
We implement a RAN simulator that supports UE registration and PDU session creation. 
In our evaluation, we generate 20K UEs where each containing 40 PDRs. Once the RAN simulator finishes UE and PDU session requests, 5GC knows how to handle both uplink and downlink packets. 
Based on TRex, we deploy scripts that can generate all kinds of traffic traces \cite{traces} (\eg with different packet sizes, flows, and patterns) to simulate traffic from UEs. 
As Figure~\ref{fig:eva_cciot_packet_size} shows, we analyze a real traffic trace, including 31M packets. 
We can observe that most of the data packets are distributed in small packets ($\sim$100 bytes) and large packets (1400 bytes$\sim$1500 bytes). 
We can conclude that large packets contribute most of the throughput. 
Therefore, unless stated otherwise, we set the packet size as 1500Byte. 

\parabf{Baselines:}
We choose two baselines: the default configuration of the commercial-version free5GC and ResQ \cite{resq}. ResQ is the first system that discovers the leaky DMA problem and solves it by reducing the number of RX descriptors. 
Although L${^2}$5GC \cite{l25gc} provides an open-source 5G core system, it still has some limitations (\eg only supports a limited number of user sessions) compared with commercial one. Therefore, our experiment uses a commercial 5G core to reveal the LLC usage problems. 

We measure the throughput, packet loss rate, RX and TX core LLC miss rates of a UPF to understand the performance of different approaches. We run each experiment multiple times to obtain reliable results reported in the paper.




\subsection{Overall performance}
\label{sec:case_study}
We compare the UPF performance when using the commercial 5G core's default configuration, ResQ \cite{resq}, and three variants of our system: (i) 5GC$^2$ache$_{12}$ stands for the variant that only adopt offline UPF configuration to alleviate the DMA leakage problem and hot/cold mbuf problem; (ii) 5GC$^2$ache$_{3}$ stands for the variant that only adopt the online LLC allocation to resolve the cache contention problem; (iii) 5GC$^2$ache$_{123}$ stands for the complete implementation.


\parabf{Throughput:} 
Figure~\ref{fig:eva_upf_case_study_throughput} shows that the UPF's throughput is 54.81Gbps when using the default configuration. The throughput increases to 58.98Gbps when using ResQ \cite{resq}. ResQ obtains similar throughput as the default configuration because the number of RX descriptors \be{derived} by ResQ is the same as the default settings, \be{which is not optimal}. 

After adopting \sysname, the throughput increases to 61.72Gbps, 71.49Gbps and 76.41Gbps when using 5GC$^2$ache$_{3}$, 5GC$^2$ache$_{12}$ and 5GC$^2$ache$_{123}$, respectively. 
Therefore, compared with the default configuration and ResQ, \sysname improves UPF's throughput by 39.41\% and 29.55\%. 

\parabf{LLC miss rate of packet reception thread:} 
Figure~\ref{fig:eva_upf_case_study_rx_llc_miss} shows the LLC miss rate of the reception thread. The LLC miss rate is 63.62\% when using the default configuration. 
This number becomes 57.29\% when using ResQ. The LLC miss rate decreases because ResQ isolates DDIO cache and core cache. 

The LLC miss rate is 46.72\%, 17.75\% and 5.68\% when using 5GC$^2$ache$_{3}$, 5GC$^2$ache$_{12}$ and 5GC$^2$ache$_{123}$, respectively. Therefore, compared with the default configuration and ResQ, \sysname decreases the LLC miss rate of the reception thread by 91.07\% and 90.09\%. 

\parabf{LLC miss rate of packet transmission thread:} 
Figure~\ref{fig:eva_upf_case_study_tx_llc_miss} shows the LLC miss rate of the transmission thread. The LLC miss rate is 30.71\% when using the default configuration. This number becomes 19.14\% when using ResQ due to isolating DDIO cache and core cache. 

One interesting observation is that the transmission core LLC miss rate is 26.05\% when using 5GC$^2$ache$_{3}$, higher than ResQ. The root cause is that the DDIO cache in 5GC$^2$ache$_{3}$ is larger than 2 ways. Therefore, the core cache in 5GC$^2$ache$_{3}$ becomes smaller and results in the increased LLC miss rate. 
However, Figure~\ref{fig:eva_upf_case_study_throughput} shows that the throughput achieved by 5GC$^2$ache$_{3}$ is higher than ResQ because 5GC$^2$ache$_{3}$ can find a sweet point among the trade-off between DDIO cache and core cache. 

The LLC miss rate of the packet transmission thread is 2.70\% and 0.50\% when using 5GC$^2$ache$_{12}$ and 5GC$^2$ache$_{123}$, respectively. 
Therefore, compared with the default configuration, \sysname decreases the LLC miss rate of the transmission thread by 98.37\%. 

{
\setlength{\abovecaptionskip}{10pt}
\begin{figure*}[t]
    \centering
    \subfigure[UPF throughput]{
    \begin{minipage}[b]{2.2in}
    \includegraphics[width=2.2in]{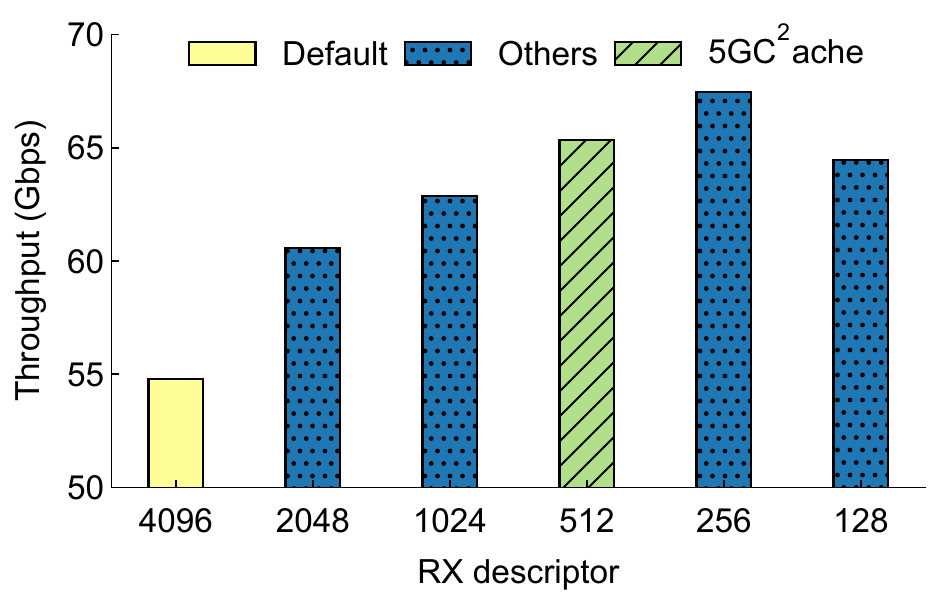}\vspace{-0.2in}
    \label{fig:eva_upf_rx_desc_throughput}
    \end{minipage}
    }
    \subfigure[UPF packet loss rate]{
    \begin{minipage}[b]{2.2in}
    \includegraphics[width=2.2in]{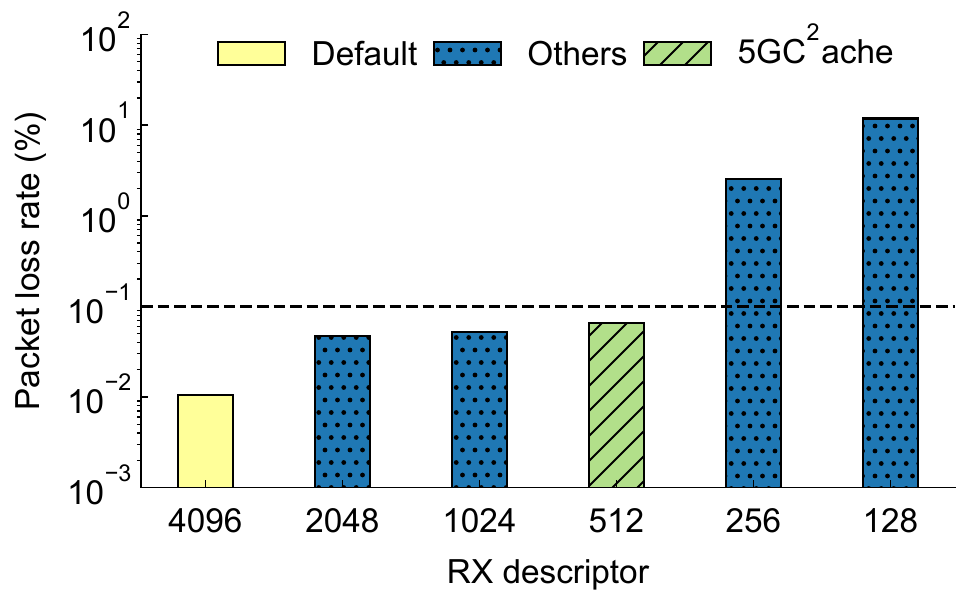}\vspace{-0.2in}
    \label{fig:eva_upf_rx_desc_pkt_loss_log}
    \end{minipage}
    }
    \subfigure[UPF TX core LLC miss rate]{
    \begin{minipage}[b]{2.2in}
    \includegraphics[width=2.2in]{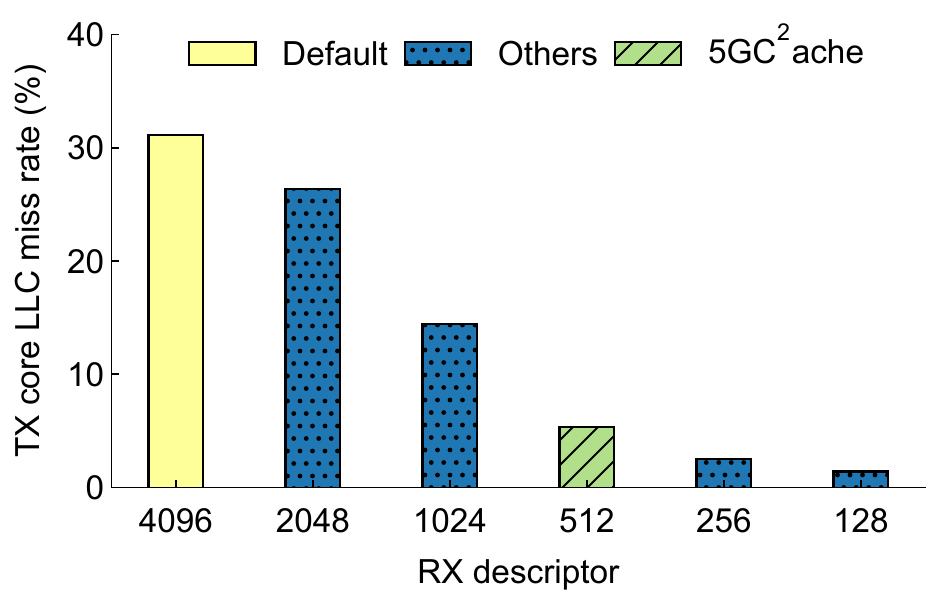}\vspace{-0.2in}
    \label{fig:eva_upf_rx_desc_tx_llcmiss}
    \end{minipage}
    }
    \caption{UPF performance under \sysname with different RX descriptors.}
    \label{fig:eva_rx_desc}
\end{figure*}
}

{
\begin{figure*}[t]
    \centering
    \subfigure[UPF throughput]{
    \begin{minipage}[b]{2.2in}
    \includegraphics[width=2.2in]{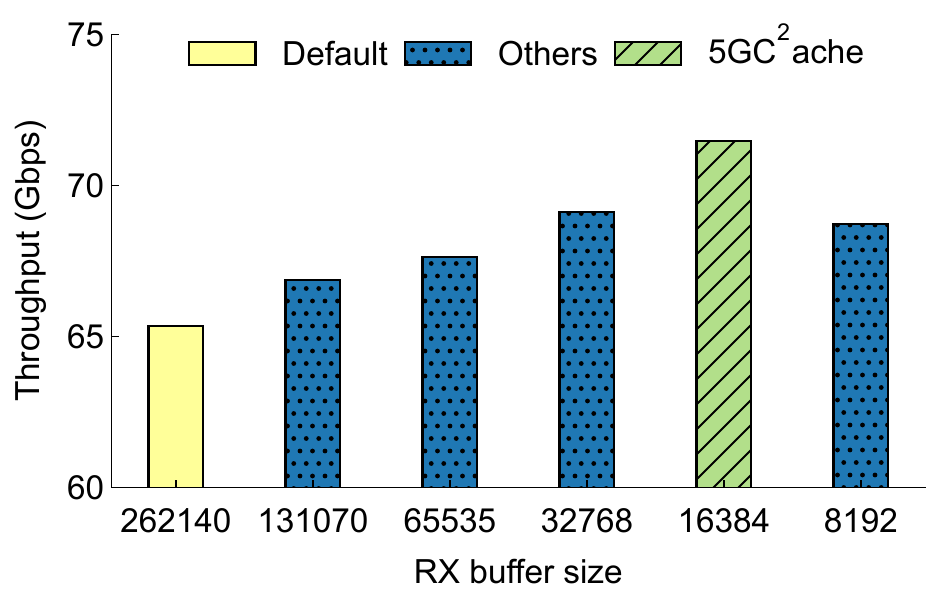}\vspace{-0.2in}
    \label{fig:eva_upf_mem_pool_throughput}
    \end{minipage}
    }
    \subfigure[UPF packet loss rate]{
    \begin{minipage}[b]{2.2in}
    \includegraphics[width=2.2in]{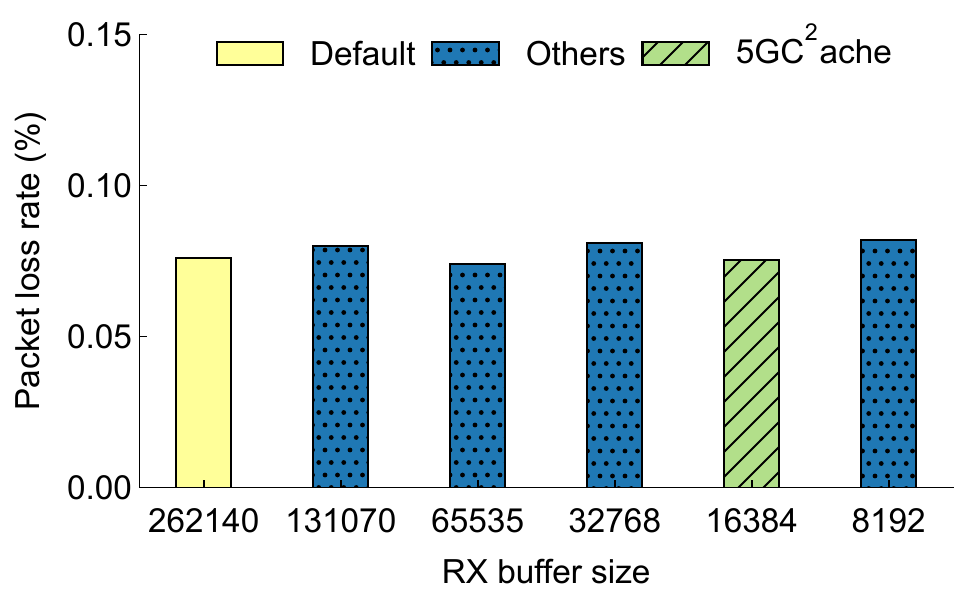}\vspace{-0.2in}
    \label{fig:eva_upf_mem_pool_packet_loss}
    \end{minipage}
    }
    \subfigure[UPF RX core LLC miss rate]{
    \begin{minipage}[b]{2.2in}
    \includegraphics[width=2.2in]{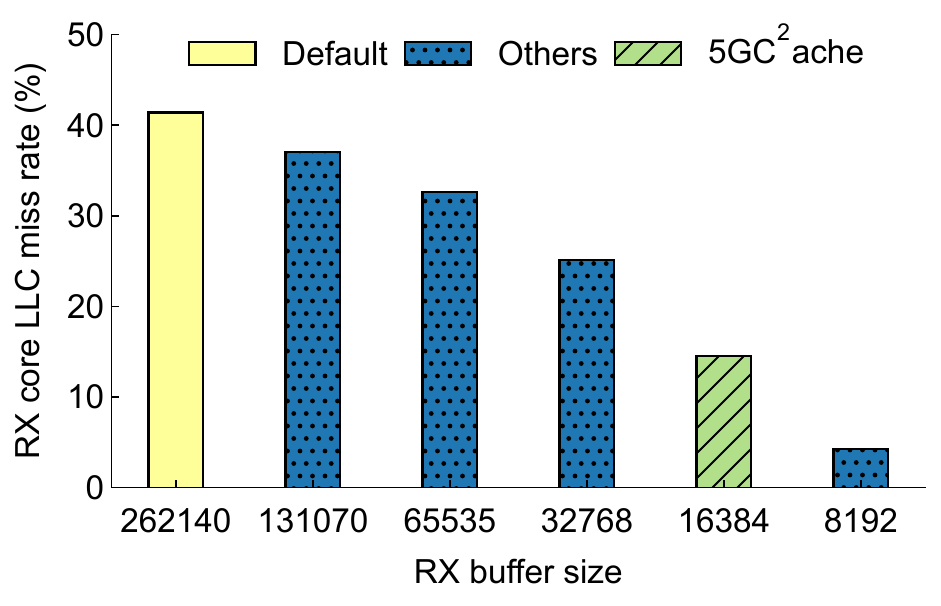}\vspace{-0.2in}
    \label{fig:eva_upf_mem_pool_rx_llcmiss}
    \end{minipage}
    }
    \caption{UPF performance under \sysname with different RX buffer sizes.}
    \label{fig:e_mem_pool}
\end{figure*}
}

\parabf{Throughput when serving traffic with different sizes of packets:} 
Figure~\ref{fig:eva_upf_case_study_diff_pkt_size} shows the UPF's throughput when the size of the ingress packet changes. We can observe that the throughput decreases when the packet size becomes smaller. When the packet size is 1500 bytes, the throughput is 54.81Gbps with the default configuration. In contrast, the throughput is 76.41Gbps when using \sysname, 39.41\% higher.

Similarly, compared to the default configuration, throughput achieved by \sysname is 40.99\%, 44.43\%, and 42.35\% larger when the packet size is 1024 bytes, 512 bytes, and 256 bytes, respectively. 
When the packet size is 128 bytes, the throughput achieved by \sysname is 36.57\% higher than the default configuration. The throughput improvement is smaller because with a smaller packet size, LLC experiences less pressure from DDIO compared to the case of a large packet size. When the packet size is 64 bytes, the throughput improvement achieved by \sysname becomes 29.66\%, the smallest in all experiments. In summary, \sysname can provide significant throughput improvement (29.66\%$\sim$44.43\%) when the 5G core serves traffic with different sizes of packets. 

\subsection{Selecting the number of RX descriptors} 
\label{sec:rx_desc}
The number of RX descriptors chosen by \sysname is 512. Figure~\ref{fig:eva_upf_rx_desc_throughput} shows that the throughput achieved by \sysname is 65.36Gbps, 19.25\% improvement than the 54.81Gbps of the default configuration.

As Figure~\ref{fig:eva_upf_rx_desc_throughput} shows, 256 RX descriptors can achieve 67.45Gbps throughput, the highest among all the cases. The 512 chosen by \sysname does not reach the highest throughput. \sysname makes this selection because it does not want to introduce a high packet loss rate.

Figure~\ref{fig:eva_upf_rx_desc_pkt_loss_log} shows that when 256 RX descriptors are used, the packet loss rate is 2.54\%, which is very high. To avoid such a high packet loss rate, \sysname sacrifices a small percentage of throughput and chooses \be{512} as its number of RX descriptors.

When using 512 RX descriptors, Figure~\ref{fig:eva_upf_rx_desc_tx_llcmiss} shows that \sysname helps the packet transmission thread to reduce LLC miss rate from 31.16\% (default configuration) to 5.32\%. Such a significant drop (82.93\%) of LLC miss rate on the packet transmission thread is caused by avoiding numerous DMA leakage described in \S~\ref{sec:leaky_dma}. 

{
\setlength{\abovecaptionskip}{3pt}
\begin{figure*}[t]
    \centering
    \subfigure[UPF throughput]{
    \begin{minipage}[b]{2.2in}
    \includegraphics[width=2.2in]{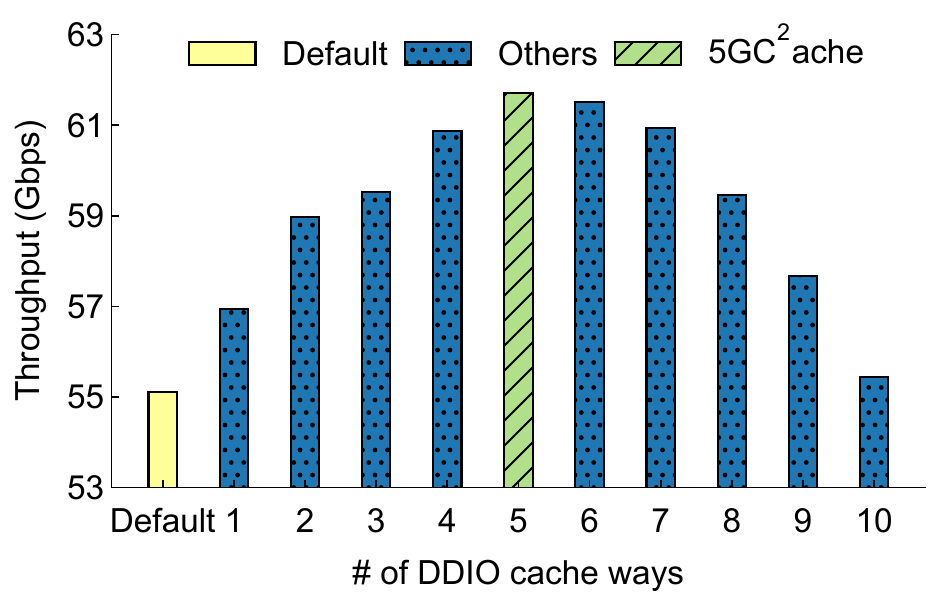}\vspace{-0.2in}
    \label{fig:eva_upf_cache_contention_throughput}
    \end{minipage}
    }
    \subfigure[UPF RX core LLC miss rate]{
    \begin{minipage}[b]{2.2in}
    \includegraphics[width=2.2in]{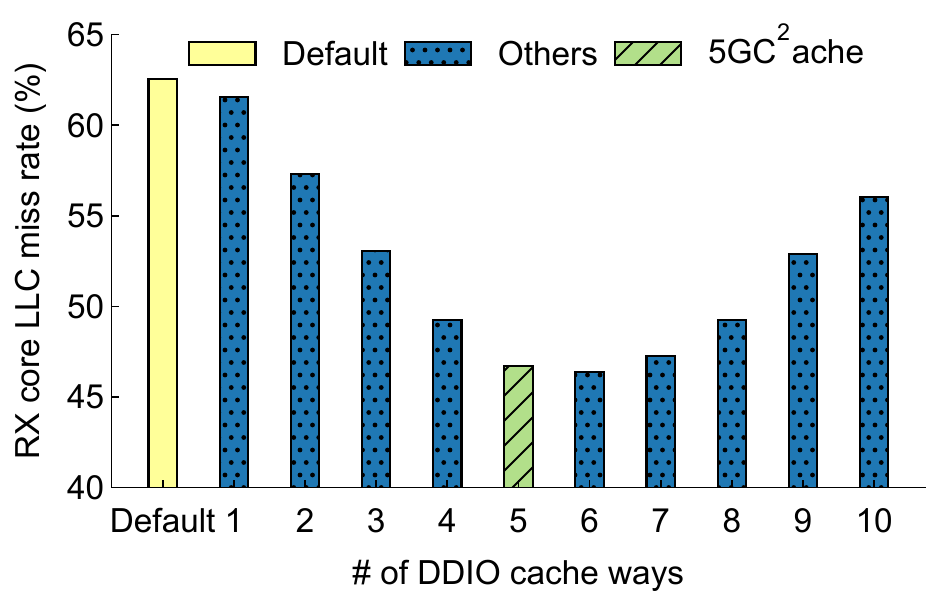}\vspace{-0.2in}
    \label{fig:eva_upf_cache_contention_rx_llcmiss}
    \end{minipage}
    }
    \subfigure[UPF TX core LLC miss rate]{
    \begin{minipage}[b]{2.2in}
    \includegraphics[width=2.2in]{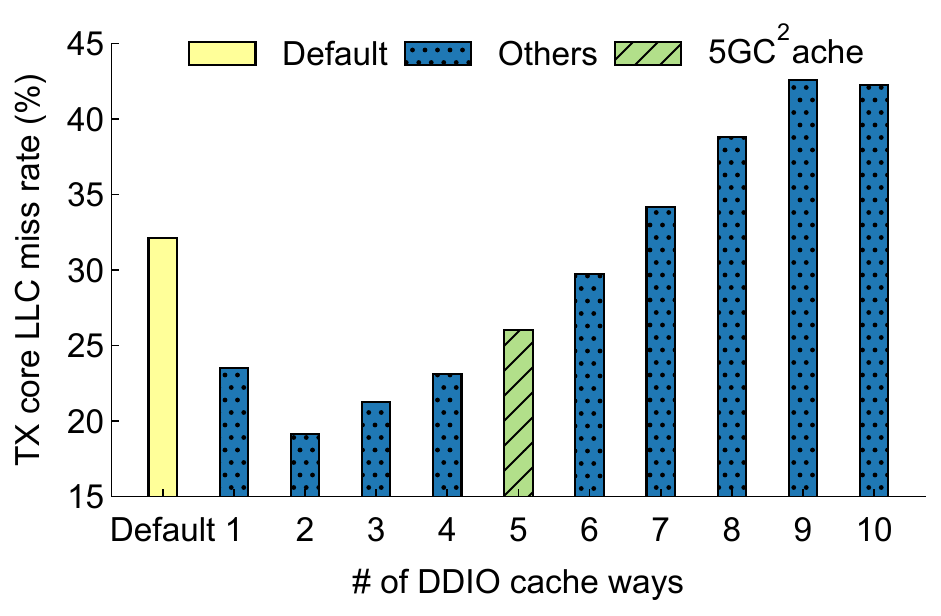}\vspace{-0.2in}
    \label{fig:eva_upf_cache_contention_tx_llcmiss}
    \end{minipage}
    }
    \caption{UPF performance under \sysname with different DDIO cache ways.}
    \label{fig:eva_cache_contention}
\end{figure*}
}

\subsection{Selecting the size of RX buffer} 
\label{sec:mem_pool}
The size of RX buffer chosen by \sysname is 16384. Figure~\ref{fig:eva_upf_mem_pool_throughput} shows that the throughput achieved by \sysname is 71.49Gbps, 9.38\% improvement than the 65.36Gbps of the default configuration.
We can also observe that RX buffer with size 16384 achieves the highest among all the cases, which is also chosen by \sysname. 

Figure~\ref{fig:eva_upf_mem_pool_packet_loss} shows that when we use the RX buffer with size 16384, the packet loss rate is 0.07\%, without significant overhead. 

When using RX buffer size as 16384, Figure~\ref{fig:eva_upf_mem_pool_rx_llcmiss} shows that \sysname helps the packet reception thread to reduce LLC miss rate from 41.40\% (default configuration) to 14.56\%. Such a significant degradation (64.83\%) of LLC miss rate on the packet reception thread is caused by resolving hot/cold mbuf issues described in \S~\ref{sec:cold_mbuf}.

\subsection{Minimizing cache contention}
\label{sec:ddio_size}
The DDIO cache ways chosen by \sysname is 5. 
Figure~\ref{fig:eva_upf_cache_contention_throughput} shows that the throughput achieved by \sysname is 61.72Gbps, 12.01\% improvement than the 55.10Gbps of the default configuration. \be{Note that \sysname can dynamically change the DDIO cache ways for different types of traffic.}

When using 5 DDIO cache ways, Figure~\ref{fig:eva_upf_cache_contention_rx_llcmiss} shows that \sysname helps the packet reception thread to reduce LLC miss rate from 62.57\% (default configuration) to 46.72\%. Such a degradation (25.33\%) of LLC miss rate on the packet reception thread is caused by reducing the cache contention described in \S~\ref{sec:cache_contention}. 

Figure~\ref{fig:eva_upf_cache_contention_tx_llcmiss} shows that
when using 5 DDIO cache ways, 
the LLC miss rate of the packet \be{transmission} thread decreases from 32.13\% to 26.05\%. We can also observe that 5 DDIO cache ways chosen by \sysname does not reach the lowest LLC miss rate. \sysname makes this decision because it needs to find a sweet point among the trade-off between DDIO cache and core cache for high performance.

\section{Related Work}
\label{sec:related}

\parabf{DDIO and LLC usage in data center:}
There are some previous works \cite{u-ddio, ddio1, ddio2} on reverse engineering the DDIO mechanism for better use LLC in the data center. 
Farshin \etal \cite{ddio1} analyze implementation details of DDIO and identify the limited LLC space allocated to DDIO.  
Wang \etal \cite{u-ddio} develop the reverse engineering analysis framework and provide a deeper understanding of DDIO. 
The above works can be used as a supplement to \sysname. 

\parabf{Leaky DMA:}
Amin \etal \cite{resq} first raise the leaky DMA problem and propose ResQ, which optimizes leaky DMA by shrinking RX descriptors to suit DDIO cache space. 
Yuan \etal \cite{ddio3} propose an IO-aware LLC allocation system to optimize leaky DMA. 
In this paper, \sysname provides an effective solution to alleviating DMA leakage for the 5G UPF. 

\parabf{Hot/cold mbuf in data center:}
Wang \etal \cite{u-ddio} analyze the buffer annealing problem and solve it by shrinking the number of RX descriptors under low traffic. However, this solution does not fit for high throughput and could cause packet loss.
Conor \etal \cite{hot_warm_cold_mbuf} propose a stack model to replace the ring model for RX buffer, aiming at optimizing mbuf usage of multi-thread applications in the data center. However, the stack model needs to modify the 5G core source code and brings locks among threads, reducing performance. 

\parabf{Cache contention:}
The cache contention problem has been widely studied in the cloud environment \cite{cache_partition1,cache_partition2,cache_partition3,cache_partition4,cache_partition5,cache_partition6}. 
Amin \etal \cite{resq} provide ResQ, addressing performance degradation by identifying and mitigating cache contention. However, static cache isolation cannot work well under different types of traffic. 
Wu \etal \cite{granularnf} propose a decomposition solution for stateful network functions to minimize the LLC contention. 
Manousis \etal \cite{slomo} propose SLOMO, which analyzes the competition for LLC, DDIO, and memory resources, without identifying the root cause of cache contention. 

\parabf{5G core systems:}
There are also several open-source 5G core systems. 
Jain \etal propose free5GC \cite{free_5gc} and L${^2}$5GC \cite{l25gc}, implementing UPF based on kernel stack and kernel bypass technologies, respectively. 
OpenAirInterface \cite{oai} focuses on RAN but also provides a 5G core. However, the above open-source 5G core systems have some limitations, such as supporting limited throughput and a limited number of user sessions. 
To the best of our knowledge, both open-source and commercial 5G core systems usually implement the 5G core without considering cache optimization in this paper.

\section{Conclusion}
\label{sec:concl}

This paper proposes \sysname, a system that adjusts the number of RX descriptors, the size of the RX buffer, and the LLC allocation between DDIO and CPU core to reduce the incorrect usage of LLC in the 5G UPF. By reducing DMA leakage, hot/cold mbuf issues, and cache contention, \sysname can significantly improve the throughput of a commercial 5G UPF from an anonymous vendor. We believe that \sysname opens up opportunities to improve 5G UPF performance from the perspective of cache optimization.

\label{lastpage}

\balance{
    {\small
    \bibliographystyle{plain}
    \bibliography{references}

\begin{thebibliography}{10}

\bibitem{23.501}
3GPP~TS 23.501.
\newblock System architecture for the 5g system;(rel-15), 2015.

\bibitem{23.502}
3GPP~TS 23.502.
\newblock Procedures for the 5g system (5gs);(rel-15), 2017.

\bibitem{29.244}
3GPP~TS 29.244.
\newblock Interface between the control plane and the user plane nodes;(rel-16), 2018.

\bibitem{33.513}
3GPP~TS 33.513.
\newblock 5g security assurance specification (scas); user plane function (upf);(rel-16), 2019.

\bibitem{llc-replace-policy}
Andreas Abel and Jan Reineke.
\newblock Reverse engineering of cache replacement policies in intel microprocessors and their evaluation.
\newblock In {\em 2014 IEEE International Symposium on Performance Analysis of Systems and Software (ISPASS)}, pages 141--142, 2014.

\bibitem{5g-background2}
Patrick~Kwadwo Agyapong, Mikio Iwamura, Dirk Staehle, Wolfgang Kiess, and Anass Benjebbour.
\newblock Design considerations for a 5g network architecture.
\newblock {\em IEEE Communications Magazine}, pages 65--75, 2014.

\bibitem{traces}
Chen Avin, Manya Ghobadi, Chen Griner, and Stefan Schmid.
\newblock On the complexity of traffic traces and implications.
\newblock SIGMETRICS '20, page 47–48, 2020.

\bibitem{descriptor}
Alexander Beifuß, Daniel Raumer, Paul Emmerich, Torsten~M. Runge, Florian Wohlfart, Bernd~E. Wolfìnger, and Georg Carle.
\newblock A study of networking software induced latency.
\newblock In {\em 2015 International Conference and Workshops on Networked Systems (NetSys)}, pages 1--8, 2015.

\bibitem{trex}
Cisco.
\newblock {Cisco TRex: Realistic Traffic Generator}.
\newblock \url{https://trex-tgn.cisco.com}.
\newblock (Accessed on 2024-01).

\bibitem{cache_partition1}
Nosayba El-Sayed, Anurag Mukkara, Po-An Tsai, Harshad Kasture, Xiaosong Ma, and Daniel Sanchez.
\newblock Kpart: A hybrid cache partitioning-sharing technique for commodity multicores.
\newblock In {\em 2018 IEEE International Symposium on High Performance Computer Architecture (HPCA)}, pages 104--117, 2018.

\bibitem{ddio1}
Alireza Farshin, Amir Roozbeh, Gerald Q.~Maguire Jr., and Dejan Kostic.
\newblock Reexamining direct cache access to optimize {I/O} intensive applications for multi-hundred-gigabit networks.
\newblock In {\em 2020 {USENIX} Annual Technical Conference, {USENIX} {ATC} 2020}, pages 673--689, 2020.

\bibitem{llc-background2}
Alireza Farshin, Amir Roozbeh, Gerald~Q Maguire~Jr, and Dejan Kosti{\'c}.
\newblock Make the most out of last level cache in intel processors.
\newblock In {\em Fourteenth EuroSys Conference (EuroSys'20)}, 2019.

\bibitem{ddio-application3}
Alireza Farshin, Amir Roozbeh, Gerald~Q Maguire~Jr, and Dejan Kostic.
\newblock Optimizing intel data direct i/o technology for multi-hundred-gigabit networks.
\newblock In {\em Fifteenth EuroSys Conference (EuroSys'20)}, 2020.

\bibitem{packetloss2}
Intel Forum.
\newblock Tuning the buffers: a practical guide to reduce or avoid packet loss in dpdk applications, 2017.

\bibitem{perf}
Linux Foundation.
\newblock {Perf: Linux profiling with performance counters}.
\newblock \url{https://perf.wiki.kernel.org/index.php/Main_Page}.
\newblock (Accessed on 2024-01).

\bibitem{free_5gc}
free5GC Organization.
\newblock {free5GC: An Open-Source Project for 5G Mobile Core Networks}.
\newblock \url{https://www.free5gc.org/}.
\newblock (Accessed on 2024-01).

\bibitem{cache_partition2}
Liran Funaro, Orna~Agmon Ben-Yehuda, and Assaf Schuster.
\newblock Ginseng: Market-driven llc allocation.
\newblock In {\em 2016 {USENIX} Annual Technical Conference {USENIX} ({ATC}'16)}, pages 295--308, 2016.

\bibitem{packetloss1}
Jim Gettys and Kathleen Nichols.
\newblock Bufferbloat: dark buffers in the internet.
\newblock {\em Communications of the ACM}, pages 57--65, 2012.

\bibitem{5g-background1}
A.~Gupta and R.~K. Jha.
\newblock A survey of 5g network: Architecture and emerging technologies.
\newblock {\em IEEE Access}, pages 1206--1232, 2015.

\bibitem{cat}
Andrew Herdrich, Edwin Verplanke, Priya Autee, Ramesh Illikkal, Chris Gianos, Ronak Singhal, and Ravi Iyer.
\newblock Cache qos: From concept to reality in the intel® xeon® processor e5-2600 v3 product family.
\newblock In {\em 2016 IEEE International Symposium on High Performance Computer Architecture (HPCA)}, pages 657--668, 2016.

\bibitem{5gc_up}
Cheng-Ying Hsieh, Yao-Wen Chang, Chien Chen, and Jyh-Cheng Chen.
\newblock Design and implementation of a generic 5g user plane function development framework.
\newblock In {\em MobiCom 2021}, page 846–848, 2021.

\bibitem{dca1}
R.~Huggahalli, R.~Iyer, and S.~Tetrick.
\newblock Direct cache access for high bandwidth network i/o.
\newblock In {\em 32nd International Symposium on Computer Architecture (ISCA'05)}, pages 50--59, 2005.

\bibitem{dpdk}
Intel.
\newblock {Data Plane Development Kit (DPDK)}.
\newblock \url{https://www.dpdk.org}.
\newblock (Accessed on 2024-01).

\bibitem{ddio}
Intel.
\newblock {Intel Data Direct I/O Technology}.
\newblock \url{https://www.intel.com/content/www/us/en/io/data-direct-i-o-technology.html}.
\newblock (Accessed on 2024-01).

\bibitem{intel_register_doc}
Intel.
\newblock Intel 64 and ia-32 architectures software developer’s manual volume 4: Model-specific registers, 2019.

\bibitem{l25gc}
Vivek Jain, Hao-Tse Chu, Shixiong Qi, Chia-An Lee, Hung-Cheng Chang, Cheng-Ying Hsieh, K.~K. Ramakrishnan, and Jyh-Cheng Chen.
\newblock L25gc: A low latency 5g core network based on high-performance nfv platforms.
\newblock In {\em SIGCOMM 2022}, page 143–157, 2022.

\bibitem{ddio-application2}
Antoine Kaufmann, SImon Peter, Naveen~Kr Sharma, Thomas Anderson, and Arvind Krishnamurthy.
\newblock High performance packet processing with flexnic.
\newblock In {\em 2016 International Conference on Architectural Support for Programming Languages and Operating Systems (ASPLOS'16)}, pages 67--81, 2016.

\bibitem{dca2}
Amit Kumar, Ram Huggahalli, and Srihari Makineni.
\newblock Characterization of direct cache access on multi-core systems and 10gbe.
\newblock In {\em 2009 IEEE 15th International Symposium on High Performance Computer Architecture (HPCA)}, pages 341--352, 2009.

\bibitem{spacecore}
Yuanjie Li, Hewu Li, Wei Liu, Lixin Liu, Yimei Chen, Jianping Wu, Qian Wu, Jun Liu, and Zeqi Lai.
\newblock A case for stateless mobile core network functions in space.
\newblock In {\em Proceedings of the ACM SIGCOMM 2022 Conference}, SIGCOMM '22, page 298–313. Association for Computing Machinery, 2022.

\bibitem{dca4}
Guangdeng Liao, Xia Znu, and Laxmi Bnuyan.
\newblock A new server i/o architecture for high speed networks.
\newblock In {\em 2011 IEEE 17th International Symposium on High Performance Computer Architecture (HPCA)}, pages 255--265, 2011.

\bibitem{slomo}
Antonis Manousis, Rahul~Anand Sharma, Vyas Sekar, and Justine Sherry.
\newblock Contention-aware performance prediction for virtualized network functions.
\newblock In {\em SIGCOMM 2020}, page 270–282, 2020.

\bibitem{ddio-application1}
Ilias Marinos, Robert~N.M. Watson, and Mark Handley.
\newblock Network stack specialization for performance.
\newblock In {\em SIGCOMM 2014}, page 175–186, 2014.

\bibitem{llc-background1}
Cl{\'e}mentine Maurice, Nicolas Le~Scouarnec, Christoph Neumann, Olivier Heen, and Aur{\'e}lien Francillon.
\newblock Reverse engineering intel last-level cache complex addressing using performance counters.
\newblock In {\em Research in Attacks, Intrusions, and Defenses: 18th International Symposium, RAID 2015}, pages 48--65, 2015.

\bibitem{cache_partition3}
Sparsh Mittal.
\newblock A survey of techniques for cache partitioning in multicore processors.
\newblock {\em ACM Computing Surveys (CSUR)}, 50:1--39, 2017.

\bibitem{Mitzenmacher_probBook_2005}
Michael Mitzenmacher and Eli Upfal.
\newblock {\em Probability and Computing: Randomized Algorithms and Probabilistic Analysis.}
\newblock 2005.

\bibitem{oai}
OpenAirInterface Organization.
\newblock {OpenAirInterface 5G Core Network Project}.
\newblock \url{https://openairinterface.org/oai-5g-core-network-project/}.
\newblock (Accessed on 2024-01).

\bibitem{leakydma2}
Amy Ousterhout, Joshua Fried, Jonathan Behrens, Adam Belay, and Hari Balakrishnan.
\newblock Shenango: Achieving high {CPU} efficiency for latency-sensitive datacenter workloads.
\newblock In {\em {NSDI} 2019}, pages 361--378, 2019.

\bibitem{localization_core}
Andrea Pinto, Giuseppe Santaromita, Claudio Fiandrino, Domenico Giustiniano, and Flavio Esposito.
\newblock Experimenting with localization management functions in 5g core networks.
\newblock In {\em MobiCom 2022}, MobiCom '22, page 806–807, 2022.

\bibitem{5gc-book}
Stefan Rommer, Peter Hedman, Magnus Olsson, Lars Frid, Shabnam Sultana, and Catherine Mulligan.
\newblock {\em 5G Core Networks: Powering Digitalization}.
\newblock Academic Press, 2019.

\bibitem{5g-background3}
Peter Rost, Albert Banchs, Ignacio Berberana, Markus Breitbach, Mark Doll, Heinz Droste, Christian Mannweiler, Miguel~A Puente, Konstantinos Samdanis, and Bessem Sayadi.
\newblock Mobile network architecture evolution toward 5g.
\newblock {\em IEEE Communications Magazine}, pages 84--91, 2016.

\bibitem{samsung_5gc}
Samsung.
\newblock {Samsung Achieves 305 Gbps on 5G UPF Core Utilizing Intel® Architecture}.
\newblock \url{https://images.samsung.com/is/content/samsung/assets/global/business/networks/insights/white-papers/1217_samsung-achieves-305gbps-on-5g-upf-core-utilizing-intel-architecture/Wireless-Core-Intel-Samsung-Performance-Whitepaper-Design-1217.pdf}.
\newblock (Accessed on 2024-01).

\bibitem{5gc_lb}
Tze-Jie Tan, Fu-Lian Weng, Wei-Ting Hu, Jyh-Cheng Chen, and Cheng-Ying Hsieh.
\newblock A reliable intelligent routing mechanism in 5g core networks.
\newblock In {\em MobiCom 2020}, 2020.

\bibitem{dca3}
Dan Tang, Yungang Bao, Weiwu Hu, and Mingyu Chen.
\newblock Dma cache: Using on-chip storage to architecturally separate i/o data from cpu data for improving i/o performance.
\newblock In {\em 2010 IEEE 16th International Symposium on High-Performance Computer Architecture (HPCA)}, pages 1--12, 2010.

\bibitem{ddio2}
Mohammadkazem Taram, Ashish Venkat, and Dean~M. Tullsen.
\newblock Packet chasing: Spying on network packets over a cache side-channel.
\newblock In {\em 47th {ACM/IEEE} Annual International Symposium on Computer Architecture ({ISCA}'20)}, pages 721--734, 2020.

\bibitem{resq}
Amin Tootoonchian, Aurojit Panda, Chang Lan, Melvin Walls, Katerina Argyraki, Sylvia Ratnasamy, and Scott Shenker.
\newblock Resq: Enabling slos in network function virtualization.
\newblock In {\em {NSDI} 2018}, page 283–297, 2018.

\bibitem{multithread_dpdk}
Conor Walsh.
\newblock Optimize memory usage in multithreaded data plane development kit (dpdk) applications, 2018.

\bibitem{hot_warm_cold_mbuf}
Conor Walsh.
\newblock Optimize memory usage in multithreaded applications, 2020.

\bibitem{u-ddio}
Minhu Wang, Mingwei Xu, and Jianping Wu.
\newblock Understanding i/o direct cache access performance for end host networking.
\newblock {\em Proceedings of the ACM on Measurement and Analysis of Computing Systems}, pages 1--37, 2022.

\bibitem{cache_partition4}
Xiaodong Wang, Shuang Chen, Jeff Setter, and José~F. Martínez.
\newblock Swap: Effective fine-grain management of shared last-level caches with minimum hardware support.
\newblock In {\em 2017 IEEE International Symposium on High Performance Computer Architecture (HPCA)}, pages 121--132, 2017.

\bibitem{pcm}
Thomas Willhalm, Roman Dementiev, and Patrick Fay.
\newblock Intel performance counter monitor - a better way to measure cpu utilization, 2017.

\bibitem{granularnf}
Ziyan Wu, Tianming Cui, Arvind Narayanan, Yang Zhang, Kangjie Lu, Antonia Zhai, and Zhi-Li Zhang.
\newblock Granularnf: Granular decomposition of stateful nfv at 100 gbps line speed and beyond.
\newblock {\em SIGMETRICS Perform. Eval. Rev.}, page 46–51, 2022.

\bibitem{cache_partition5}
Yaocheng Xiang, Xiaolin Wang, Zihui Huang, Zeyu Wang, Yingwei Luo, and Zhenlin Wang.
\newblock Dcaps: Dynamic cache allocation with partial sharing.
\newblock In {\em Proceedings of the Thirteenth EuroSys Conference}, EuroSys '18, 2018.

\bibitem{cache_partition6}
Cong Xu, Karthick Rajamani, Alexandre Ferreira, Wesley Felter, Juan Rubio, and Yang Li.
\newblock Dcat: Dynamic cache management for efficient, performance-sensitive infrastructure-as-a-service.
\newblock In {\em Proceedings of the Thirteenth EuroSys Conference}, EuroSys '18, 2018.

\bibitem{5g-background4}
Volkan Yazıcı, Ulas~C. Kozat, and M.~Oguz Sunay.
\newblock A new control plane for 5g network architecture with a case study on unified handoff, mobility, and routing management.
\newblock {\em IEEE Communications Magazine}, pages 76--85, 2014.

\bibitem{ddio3}
Yifan Yuan, Mohammad Alian, Yipeng Wang, Ren Wang, Ilia Kurakin, Charlie Tai, and Nam~Sung Kim.
\newblock Don't forget the {I/O} when allocating your {LLC}.
\newblock In {\em 48th {ACM/IEEE} Annual International Symposium on Computer Architecture, {ISCA} 2021}, pages 112--125, 2021.

\bibitem{zte_5gc}
ZTE.
\newblock {ZTE 5G Core Network}.
\newblock \url{https://sdnfv.zte.com.cn/en/products/VNF/5G-core-network}.
\newblock (Accessed on 2024-01).

\end{thebibliography}
    }
    \appendix

\section{DMA leakage ratio analysis}
\label{apx:hash_collision}

A quantitative analysis of the level of DMA leakage can be applied via a ``balls into bins'' model. Suppose that $N$ balls (packets) are thrown sequentially into $M$ bins (cache lines) with each bin chosen independently and uniformly at random. To help clarify the following reasoning, we call the first ball in a given bin the baseball and the rest in that bin the extra balls. Note that after throwing all the $N$ balls, the follow 3 statements will always hold,

\begin{itemize}
\item The number of baseballs is equal to the number of non-empty bins.\label{state.1}
\item The number of total DMA leakage times is equal to the number of total extra balls.\label{state.2}
\item The sum of the base and extra balls is equal to the total balls, which is $N$.\label{state.3}
\end{itemize}

Let $\boldsymbol{X}$, $\hat{\boldsymbol{X}}$ and $\boldsymbol{Y}$ be random variables representing the number of non-empty bins, empty bins, and DMA leakage times, respectively. Then by statement \ref{state.1} $\boldsymbol{X}$ is equal to the base balls and by statement \ref{state.2} $\boldsymbol{Y}$ is equal to the extra balls, thus by statement \ref{state.3}
\begin{align}
\boldsymbol{Y}+\boldsymbol{X}=N \label{eq_hash_1}
\end{align}
Recall that $M$ is the number of total bins, thus
\begin{align}
M=\boldsymbol{X}+\hat{\boldsymbol{X}}
\end{align}
based on which \eqref{eq_hash_1} can be written as,
\begin{align}
\boldsymbol{Y}=N-M+\hat{\boldsymbol{X}}
\end{align}
and the expected value of $\boldsymbol{Y}$ is,
\begin{align}
\mathbb{E}(\boldsymbol{Y})=N-M+\mathbb{E}(\hat{\boldsymbol{X}})
\end{align}
We can leverage existing results on bloom filters to calculate $\mathbb{E}(\hat{\boldsymbol{X}})$. By the equation on page 111 of \cite{Mitzenmacher_probBook_2005}, the expected fraction of empty bins is,
\begin{align}
p'=\left (1-\frac{1}{M}\right )^{N}
\end{align}
which gives,
\begin{align}
\mathbb{E}(\hat{\boldsymbol{X}})=Mp'=M\left (1-\frac{1}{M}\right )^{N}
\end{align}
Therefore,
\begin{align}
\mathbb{E}(\boldsymbol{Y})=N-M+M\left (1-\frac{1}{M}\right )^{N}
\end{align}
Further, by the last equation on page 111 of \cite{Mitzenmacher_probBook_2005}, 
\begin{align}
\Pr(|\hat{\boldsymbol{X}}-\mathbb{E}(\hat{\boldsymbol{X}})|\geq \epsilon M)\leq 2e\sqrt{M}e^{-M\epsilon^2/3p'}
\end{align}
which means that $\hat{\boldsymbol{X}}$ is strongly concentrated around its expected value, and consequently $\boldsymbol{Y}$ is also strongly concentrated around $\mathbb{E}(\boldsymbol{Y})$.

}

\end{document}